\documentclass[11pt]{article} 

\usepackage[utf8]{inputenc} 

\usepackage{geometry} 
\usepackage{graphicx} 
\usepackage{xcolor}
\usepackage{parskip}
\usepackage{doi}

\usepackage{apacite}
\usepackage{booktabs}
\usepackage{multirow}
\usepackage{color}
\usepackage{amsmath,amssymb,amsthm}
\usepackage{soul}
\usepackage{mathtools}

\newcommand{\dd}{\mathrm{d}}
\newcommand{\Sgn}{\operatorname{Sgn}}

\usepackage{hyperref}
\usepackage{bm}
\usepackage{amsmath}
\usepackage{amssymb}
\usepackage{subcaption}
\usepackage{footnpag}
\usepackage{natbib}
\setcitestyle{authoryear,citesep={; },notesep={; },round,aysep={,},yysep={;}}
\usepackage{indentfirst}
\usepackage{mathrsfs}
\usepackage{lineno}

\title{Exploring tidal dissipation in rubble pile binary secondaries using a discrete element model}
\author{Ethan R. Burnett\thanks{Assistant Professor, Department of Aerospace Engineering and Engineering Mechanics, University of Texas at Austin, 2617 Wichita St., Austin, TX 78712, U.S.A. \texttt{ethan.burnett@utexas.edu}}, \ Iosto Fodde\thanks{Assistant Professor, Department of Aerospace Science and Technology, Politecnico di Milano, Via La Masa 34, 20156 Milan, Italy}, John Wimarsson\thanks{Postdoctoral Researcher, Space Research \& Planetary Sciences, Physics Institute, University of Bern, Gesellschaftsstrasse 6, Bern, 3012, Switzerland}, Eric Frizzell\thanks{Marie Skłodowska-Curie Postdoctoral Fellow, Department of Aerospace Science and Technology, Politecnico di Milano, Via La Masa 34, 20156 Milan, Italy}, and Fabio Ferrari\thanks{Associate Professor, Department of Aerospace Science and Technology, Politecnico di Milano, Via La Masa 34, 20156 Milan, Italy}}

\date{}
     
\begin{document}
\maketitle
\begin{abstract}
In this work, models of rubble pile binary secondaries are simulated in various spin states in a system similar in size and scale to Didymos-Dimorphos. The numerical modeling is performed in the N-body Chrono-based software GRAINS, which simulates gravity, contact, and friction forces acting on non-spherical mass elements. In a high-fidelity, high-performance computational setting, tidal dissipation successfully emerges in the simulations as a consequence of the effects of inter-element contact and friction across thousands of simulated nonspherical mass elements. We devise computational techniques and rigorous numerical procedures for studying tides in the well-established framework of GRAINS. 
The simulations reveal tidal deformation, phase lag, and secular spin-down, including the transition from super-synchronous rotation to synchronous libration. We also observe the exchange of energy and angular momentum between the secondary's rotation and orbit, and identify the semidiurnal tidal response through
spectral and topographical analyses. Using both energy and torque-based calculations, we compute values of $Q/k_{2} \sim\mathcal{O}(10^{2})$, whereas this value has been predicted by other studies to lie in the range $10^{2}<Q/k_{2}<10^{6}$. 
Further study is required to determine  whether this result is representative of truly enhanced dissipativity, or a consequence of material or modeling assumptions.
 More broadly, this framework provides a numerical laboratory for investigating the granular mechanisms underlying tidal dissipation and their relationship to the macroscopic tidal response of rubble-pile bodies.
\end{abstract}

\section{Introduction}
Asteroids are abundant small primordial rocky worlds, leftovers from the earliest days of the solar system. Understanding their composition, morphology, and physical properties is key to understanding the origin and evolution of the rocky planets. Some large asteroids regularly cross paths with Earth. Due to the long-term risk of collision, improved understanding of their dynamical evolution could also even be necessary for ensuring the long-term survival of human civilization. In its upcoming 2029 close encounter with Earth, the $\sim$375 m-wide asteroid (99942) Apophis will dip to within about 5 Earth radii of the Earth's surface (significantly closer than the Moon), and will be visible to the naked eye, serving as a dramatic example of the need to track these objects and accurately predict their future paths. Multiple forces govern their dynamical evolution, beyond the familiar gravitational pull of the sun and planets. The Yarkovsky–O'Keefe–Radzievskii–Paddack (``YORP") effect \citep{RubincamYORP,Yarkovsky2002}, a thermal radiation effect, affects spin and orbital evolution via unbalanced torque and force. These bodies are also strongly affected by tidal forces, which can occur during close encounters with planets. Such encounters can adjust their spin states, altering the subsequent YORP forces, leading to changes in the orbit. The tidal response of these bodies, which determines the net effect of the encounter on the body's spin, is poorly understood. Unlike planets and moons, these bodies are ``rubble piles", highly fractured into individual boulders held together by gravity and perhaps other forces, as opposed to the continuous rheology of larger bodies. There is no complete theory governing the tidal response of such bodies.

\begin{figure}[h!]
\centering
\includegraphics[scale=1.2]{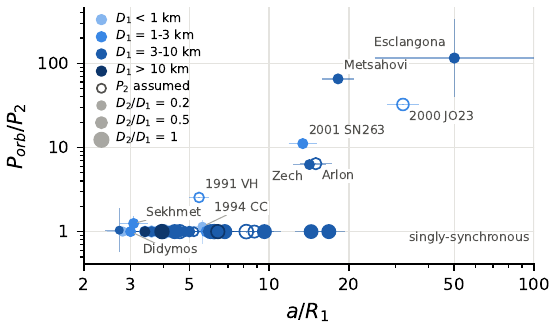}
\caption{Observed spin states (orbital to secondary spin ratio as $P_{orb}/P_2$) of binary asteroid secondaries as a function of the semi-major axis $a$ in units of primary radii $R_1$\citep{pravecDatabase}. Most secondaries are in a singly-synchronous state, whereas for wider binaries this is not necessarily the case.}
\label{fig:binarySpinStates}
\end{figure}

Interestingly, $\sim$15\% of asteroids are binary pairs \citep{Pravec2006}, providing natural laboratories for understanding asteroid dynamics. In many cases, the primary is spinning fairly quickly, and leading theories suggest that the smaller secondaries are formed from the material of the primary via binary fission \citep{Scheeres_fission} or by re-accretion following a mass-shedding event \citep{Walsh2012,Wimarsson_GRAINS24}. Binary systems are an arena for the interaction of many complex natural forces. There is the gravitational attraction between the two bodies, which are often separated by just a few radii, so non-spherical shape effects are important, introducing strong orbit-attitude coupling effects \citep{Hou_2017}. Additionally the YORP effect is still important \citep{Holsapple_YORP}, along with the so-called binary YORP (``BYORP"), which can affect the orbital evolution of the secondary, driving it further from, or even closer to, the primary \citep{Cuk_BYORP}. Most of the time these binary systems are ``singly-synchronous" -- the secondary's spin period matches its orbital period, thus always showing the same face towards the primary \citep{Pravec_SecondaryRotationSurvey}. This suggests that the process by which spin energy is dissipated in binary secondaries tends to take place very quickly in comparison to the lifetime of these systems. Even in the case of the comparatively uncommon asynchronous binaries, a lot of the time they are widely separated, the end result of what is thought to be a complex chain of BYORP then YORP-driven dynamical evolution initiated by early tidal locking \citep{Jacobson_2014}. Fig. \ref{fig:binarySpinStates} shows the spin state of binary secondaries\footnote{Data taken from the following database: \url{http://space.asu.cas.cz/~asteroid/binastdata.htm}, accessed on 19-10-2026.}, which show that most secondaries are in the singly-synchronous state, with most of the systems not in this state being wide binaries with large semi-major axes (tidal torque strength drops off as $a^{-6}$, see section \ref{subsec:theory}, thus this is as expected). Nonetheless there are also a few ``tight" asynchronous binaries, with relatively small separation of primary and secondary. In these rare cases, their shape, composition, or recent encounters could have played a role in the unsettled spin states, but it is not generally known.

This work discloses results from numerical studies using a discrete element model to explore tidal dynamics in binary asteroid systems, as well as in flyby scenarios \citep{Burnett_EGU24,Burnett_EPSC24}. The goal of this work is to develop tools and techniques to quantitatively explore, in high fidelity, the nature of dissipation of rubble pile asteroids. We apply this first to binary secondaries in binary asteroid systems. There are many past works that are relevant to this. \cite{Pravec_SecondaryRotationSurvey} gives a nice survey of binary secondary rotation states. \cite{SSSBs} reviews the numerical and experimental research carried out recently in an effort to study rubble pile asteroids. We also refer the reader to the basics of viscoelastic tidal theory as discussed in \cite{SSD_1999} and references therein. While binary asteroid dynamics are governed by granular mechanics, the de-facto prevailing assumption in the scientific community is that in bulk, these bodies can be approximated by the same first-order viscoelastic tidal theory used to describe the tidal evolution of planets and moons. 

Along this line of thought, \cite{Taylor_2011} develop a tidal evolution model for binary asteroid systems, allowing them to assess conditions for spin synchronization and binary stability. \cite{NimmoMatsuyama} and \cite{GoldreichSari} derive, from first principals and order-of-magnitude analysis, effective quality factors $Q$ and rigidity $\mu$ for rubble pile asteroids. \cite{GoldreichSari} also predicts a proportional relationship between tidal love number $k_{2}$ and asteroid mean radius, whereas \cite{Jacobson_2011} shows that a best-fit to astronomical data predicts an inversely proportional relationship. \cite{Meyer_EnergyDissBinary} performs a numerical study of the coupled orbital and attitude dynamics of a rigid ellipsoidal Dimorphos orbiting about a spherical Didymos, with imposed tidal dissipation via a generalized, non-planar MacDonald-like tidal torque term acting on the rotational state of Dimorphos, and a MacDonald tidal torque on Didymos in turn affecting the orbit of the secondary. Those simulations of 12-state nonlinear dynamics are studied on timescales of $10^{2}$ yrs., revealing long-term changes to the attitude of Dimorphos and its orbit about Didymos. They also note that for some imposed values of $Q/k_{2}$, noticeable reduction in libration amplitude occurred on a timescale of years, with relevance to the upcoming Hera survey of Didymos. 

In the realm of high-fidelity numerical modeling, \cite{TidalNBody} constructs an $N$-body damped ``spring-mass" model of a planet to explore fundamentals of viscoelastic tidal theory, also discussing how the model is to be tuned to emulate realistic rheologies. Recently, \cite{DellaGiustina24} simulated a rubble pile binary system using a discrete element model, and noted the emergence of time-varying tidal strain in the primary, computing also a tidal quality factor $Q$, as part of a longer work demonstrating the detectability of tidal distortions on the primary body using modern seismometers. That study was conducted using the $N$-body collisional code pkdgrav \citep{Richardson_pkdgrav}, based on 1500 spherical constituent particles for the primary and 500 for the secondary. Our work is conducted in GRAINS, a Chrono-based $N$-body simulation architecture, simulating nonspherical mass elements, that has successfully been used for past studies of rubble pile asteroid morphology and dynamics \citep{Fabio_Nbody,Ferrari_2020,Chrono_conference2016}. Note in this work we use the terms ``$N$-body" and ``discrete element" synonymously, but when discussing past work we use the authors' own preferred terminology.

This work appears on the eve of the arrival of the ESA Hera mission to the Didymos-Dimorphos binary asteroid system. On September 26, 2022, the DART spacecraft hit the smaller secondary, Dimorphos, at a speed of 6.6 km/s \citep{Daly_2024}, the immediate aftermath of which was observed by both the nearby Italian LICIACube spacecraft and also by ground-based telescopes. The impact was sufficient to generate a reduction of Dimorphos' originally $\sim$12 hr orbit period by about 33 minutes \citep{DART_OrbitChange_Nature}, and the resulting momentum deflection efficiency, ejecta cone geometry, and ejected mass provided a lot of useful information, such as constraints on the bulk strength and density of Dimorphos \citep{Dimoprhos_from_DART}. \cite{Agrusa_2021} studied the types of attitude instabilities that could result from the DART impact, and \cite{Dimoprhos_from_DART} noted that based on simulations, the DART impact likely caused global deformation and resurfacing of Dimorphos. \cite{Naidu_2024} detailed post-impact changes to Dimorphos' orbit about Didymos, as well as post-impact measurements of an elongated ellipsoidal shape which departed considerably from its pre-impact shape. Recently \cite{Dimorphos_LightCurve}  was able to make some constraints on the post-DART spin state of Dimorphos via lightcurve analysis. The Hera mission will provide further information about the reshaping from DART, the changes to Dimorphos' spin  \citep{Michel_2022}, and from this, there could be some constraints on the dissipativity of Dimorphos. 

The novel contribution of this work is to provide numerical recipes for studying tidal dynamics and dissipation in binary rubble pile systems via numerical simulation in discrete element routines. We apply this to disturbed secondaries in spin states of super-synchronous spin (i.e., an object spins faster than it revolves) or non-principal axis rotation (i.e., an object is in a different spin state from the classically stable spin about the maximum principal axis). These cases have relevance to the early history of such bodies before their tidal locking, or the aftermath of collisions such as that of the DART mission. The work is conducted in GRAINS, which has some advantages over traditional spherical particle-based methods, including the ability to capture interlocking effects which result from the non-sphericity of the constituent particles \citep{Fabio_Nbody,Ferrari_2020}. The focus on the secondary allows us to achieve as small a particle size as possible for a given particle ``budget". Furthermore it affords the benefit of being able to infer $Q/k_{2}$ from the evolution of attitude dynamics, as opposed to the slower and more complex orbital dynamics. This is necessary because  numerical overhead presently precludes long-term orbital mechanics studies using GRAINS alone. We successfully directly observe the tidal bulge traversing the surface of the binary secondary, and furthermore we characterize the temporal lag due to dissipativity. We also directly recover the resulting tidal torques and use them to obtain, to our knowledge, the first numerical estimate of the effective tidal dissipativity $Q/k_{2}$ of a rubble-pile secondary in a discrete-element model of a binary system. This gauge of dissipativity is computed via two independent and complementary methods: one based on angular momentum, and one based on system energy.

\section{Methods}
\subsection{The GRAINS Model}
\begin{figure}[h!]
\centering
\includegraphics[]{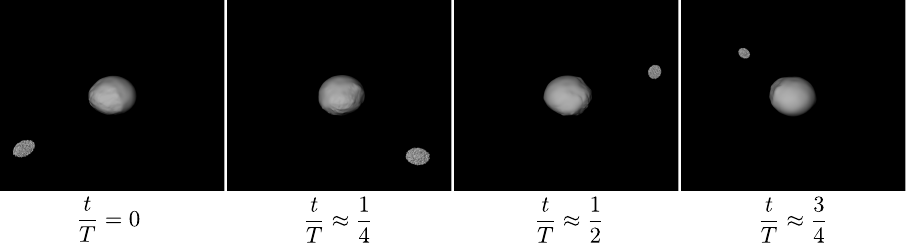}
\caption{A binary asteroid simulation in GRAINS. The spin state of the secondary is initialized with sufficient angular velocity to yield bounded but significant yaw from the minimum-energy attitude (where the long axis is collinear with the orbit radial direction). The notation $t/T\approx \frac{1}{4}$ means that roughly one-fourth of an orbit is completed.}
\label{fig:vidGRAINS}
\end{figure}
The simulations performed for this work are conducted in GRAINS, a Chrono-based $N$-body simulation architecture developed at the Politecnico di Milano. It is a discrete element modeling (DEM) approach. GRAINS models gravitational, contact, and friction dynamics between discretized non-spherical mass elements. Previously it has been used to study the aggregation of rubble piles from boulder fields \citep{Fabio_Nbody,Ferrari_2020}, formation of binary secondaries after rotational failure \citep{Wimarsson_GRAINS24,Wimarsson_25}, shape equilibrium of rubble piles \citep{Fabio_IcarusTop}, and the dynamical evolution of the Didymos-Dimorphos binary asteroid system post-impact \citep{Agrusa_2022,Ferrari_PSJ2022} There are also a number of other studies of rubble pile asteroid systems currently ongoing. For this work, each mass element in the simulation is modeled as a convex body of a maximum of 16 vertices and an average of about 10. All mass elements are of approximately the same size but differ in shape. The setup of the rubble pile secondary is similar to the GRAINS setup in \cite{Agrusa_2022}. 

In this work, we simulate the motion of a rubble pile secondary in binary asteroid systems similar to the Dimorphos-Didymos system in size and scale, but the secondary is initialized in highly non-equilibrium spin states to induce and study more pronounced dynamic tidal effects in young, unsettled (or otherwise perturbed) binary asteroid systems. We use the pre-impact shape of Dimorphos for lack of a good post-impact model. 
Due to runtime considerations and a focus on the attitude dynamics of the secondary, the primary is modeled with point-mass gravity and the numerical budget of discrete mass elements is devoted to modeling the secondary. Details of the model and various considerations are discussed as needed in the sections that follow. A visual depiction of one orbit from an example GRAINS simulation of the binary system is provided in Fig.~\ref{fig:vidGRAINS}. In our studies we model rubble pile secondaries in a range of $\sim$1700-3000 elements, and this particular model is on the low end of that range. The ``$t/T$" notation, used frequently in this work, is a non-dimensional representation of time in terms of Dimorphos orbit periods.

\subsection{Building a numerical tidal physics ``sandbox"}
\subsubsection{Orientation and frames}
\begin{figure}[h!]
\centering
\includegraphics[scale=0.8333]{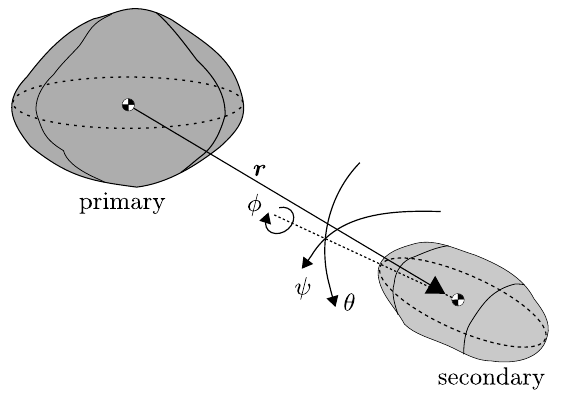}
\caption{Asteroid orientation showing the Euler angle sequence $\psi, \theta, \phi$ describing the 3-2-1 rotation from orbit frame $\mathcal{O}'$ to a body frame $\mathcal{B}$.}
\label{fig:AsteroidBFrame}
\end{figure}

We start by defining the orbit frame $\mathcal{O}' =\left\{-\hat{\bm{e}}_{r}, -\hat{\bm{e}}_{t}, \hat{\bm{e}}_{h}\right\}$, where $\hat{\bm{e}}_{r} = \bm{r}/r$ is the unit vector pointing from primary to secondary centers of mass, $\hat{\bm{e}}_{h}$ points in the orbital angular momentum direction, and $\hat{\bm{e}}_{t}$ completes the right-handed triad and points roughly along the orbital velocity direction. Next, the principal axis frame $\mathcal{B}$ diagonalizes the inertia tensor into $\text{diag}(A,B,C)$ for principal moments of inertia (MoI) $C\geq B \geq A$. We denote rotations from one reference frame ``$\mathcal{A}$" to another ``$\mathcal{B}$" as $[\mathcal{BA}]$. Where needed, vectors are resolved in a particular frame, e.g. $^{\mathcal{B}}\bm{r}$. In this work we make use of a yaw-pitch-roll attitude formulation with respective angles $\psi, \theta, \phi$ defined below in terms of components of the rotation matrix $[\mathcal{BO}']$:
\begin{subequations}
\label{YawPitchRoll}
\begin{align}
\psi = & \ \text{atan2}\left( \mathcal{BO}'_{1,2}, \mathcal{BO}'_{1,1} \right) \\
\theta = & \sin^{-1}{\left( \mathcal{BO}'_{1,3} \right)} \\
\phi = & \ \text{atan2}\left( \mathcal{BO}'_{2,3}, \mathcal{BO}'_{3,3} \right)
\end{align}
\end{subequations}
Thus $\psi = \theta = \phi = 0$ denotes the case of the long axis pointing towards the primary, with all three principal axes parallel to a basis vector of $\mathcal{O}'$. The body frame definition is depicted in Fig.~\ref{fig:AsteroidBFrame}. This choice renders counter-clockwise super-synchronous rotation to have positive sign $\dot{\psi}>0$ for consistency with foundational literature on spin and tides \citep{SSD_1999, Goldreich_Peale_SpinOrb}. We also define an arbitrary inertial frame $\mathcal{N}$, fixed with respect to the stars, in which the calculations of GRAINS are generally conducted. 

Because the rubble pile secondary is not a rigid body, the topography is not fixed in the principal axis frame $\mathcal{B}$. For simulations with super-synchronous rotation, we observe some longitudinal oscillation of the topography in this frame. To facilitate easy topographical studies, an alternate body frame definition is now discussed. Noting that the deep interior of the asteroid has greatly reduced rock movement in comparison to the shallower depths, due to increased overburden pressure \citep{NimmoMatsuyama}, a ``topography-fixed" frame $\mathcal{G}$ can be defined, in which the location of select rocks deep in the interior is fixed by definition. Being careful to exclude any of the few deep interior mass elements that undergo appreciable migration or settling, a basis is constructed of the position of at least three satisfactory elements via the following process. First, the coefficient vectors are computed:
\begin{equation}
\label{TopoF1}
W_{\star} = \begin{bmatrix} ^{\mathcal{N}}\delta\hat{\bm{r}}^{1}(t_{\star}) & ^{\mathcal{N}}\delta\hat{\bm{r}}^{2}(t_{\star}) & ^{\mathcal{N}}\delta\hat{\bm{r}}^{3}(t_{\star}) \end{bmatrix}
\end{equation}
\begin{subequations}
\label{TopoF2}
\begin{align}
\bm{\alpha} = & \ W_{e}^{-1}[\mathcal{BN}_{e}]^{\top}\hat{\bm{e}}_{1} \\
\bm{\beta} = & \ W_{e}^{-1}[\mathcal{BN}_{e}]^{\top}\hat{\bm{e}}_{2} \\
\bm{\gamma} = & \ W_{e}^{-1}[\mathcal{BN}_{e}]^{\top}\hat{\bm{e}}_{3} 
\end{align}
\end{subequations}
where in this context $\hat{\cdot}$ denotes a vector of unit norm, the $\hat{\bm{e}}_{i}$ are basis vectors e.g. $\hat{\bm{e}}_{1} = (1, 0, 0)^{\top}$, and otherwise subscripts $\left( \cdot \right)_{\star} = \left(\cdot \right)(t_{\star})$ indicate evaluation of a quantity at a specified time $t_{\star}$, the time $t_{e}$ is specifically the \textit{epoch time}, and $\delta\hat{\bm{r}}^{i} = (\bm{r}^{i} - \bm{r})/(\|\bm{r}^{i} - \bm{r}\|)$ is the vector from the barycenter of the secondary to the $i$\textsuperscript{th} element. Note that elsewhere in this paper mass element indexing is typically associated with \textit{subscripts}. Generally in this study we choose rocks at a depth of $\|\delta\bm{r}^{i}\| \leq 0.4R$. 
Having defined the basis vectors, the rotation can be resolved at any time $t_{k}\geq t_{e}$ as below:
\begin{subequations}
\label{TopoF3}
\begin{align}
\bm{g}^{1}(t_{k}) = & \ W_{k}\bm{\alpha} \\
\bm{g}^{2}(t_{k}) = & \ W_{k}\bm{\beta} \\
\bm{g}^{3}(t_{k}) = & \ W_{k}\bm{\gamma} 
\end{align}
\end{subequations}
\begin{equation}
\label{TopoF4}
[\mathcal{GN}]_{k} = \begin{bmatrix} \hat{\bm{g}}^{1}(t_{k}) & \hat{\bm{g}}^{2}(t_{k}) & \hat{\bm{g}}^{3}(t_{k})  \end{bmatrix}
\end{equation}
\begin{equation}
\label{TopoF5}
[\mathcal{GO}'] = [\mathcal{GN}][\mathcal{ON}]^{\top}[\mathcal{OO}']
\end{equation}
\begin{equation}
\label{TopoF6}
[\mathcal{OO}'] = \begin{bmatrix} -1 & 0 & 0 \\ 0 & -1 & 0 \\ 0 & 0 & 1 \end{bmatrix}
\end{equation}

Via the computational procedure of Eqs.~\eqref{TopoF1} - \eqref{TopoF6}, the epoch rotation matrices $[\mathcal{GO}']_{e}$ and $[\mathcal{BO}']_{e}$ are the same, and the $\mathcal{B}$ and $\mathcal{G}$ frames are aligned at time $t_{e}$. 
These calculations, and really any calculations of interest, should be performed \textit{after} any initial short transient shifting and settling behavior, i.e. we require $t_{e} > t_{0}$. Note explicitly that any significant relative movements of any of the rocks used to compute the basis vectors in Eq.~\eqref{TopoF4} will cause the matrix $[\mathcal{GN}]$ to become non-orthonormal. This can be addressed, if needed, by adopting a procedure to ensure preservation of orthonormality, and perhaps improved by computing the direction cosine matrix using more than 3 vector directions (somewhat related to Wahba's problem, \cite{Shuster_Wahba}). The nearest orthonormal rotation matrix $[\mathcal{GN}]_{k}$ to a non-orthonormal $[\tilde{\mathcal{GN}}]_{k}$ can be obtained via singular value decomposition:
\begin{subequations}
\label{SVD_rot}
\begin{align}
[\tilde{\mathcal{GN}}]_{k} & = U_{k}\Sigma_{k} V_{k}^{\top} \\
[\mathcal{GN}]_{k} & = U_{k}V_{k}^{\top}
\end{align}
\end{subequations}
\subsubsection{Initialization of the Rubble Pile}
For this work, rubble pile models are constructed via the same aggregation procedure as prior GRAINS works, so the interested reader is deferred there \citep{Fabio_Nbody,Ferrari_2020}. Initial transient behavior occurs in our simulations whenever the initial conditions of a given rotational scenario are different from the conditions that generated the original rubble pile aggregate. The differential spin disrupts the rubble pile secondary slightly from its old equilibrium, resulting in a rapid re-settling of the constituent particles. This manifests as a small ``jump" in the eigenvalues of the inertia tensor, which furthermore manifests in some other plots as well. In our studies we often reuse the same model in different rotational scenarios, so the initial transience is unavoidable. Otherwise, the analyses here avoid the initial set of orbits to avoid including these transients in the results. The specific selection of the amount of initial orbits to ignore is discussed in Section \ref{sec:results}.

\subsubsection{Contact Modeling}
Contact is modeled using the Smooth Contact Method of Chrono. Contact is emulated via a viscoelastic model, resolving contact-normal and tangential forces as below \citep{Chrono_conference2016}:
\begin{subequations}
\label{FrictionForce}
\begin{align}
\bm{F}_{f,n} = & \ K_{n}\delta_{n}^{p}\hat{\bm{n}} - \gamma_{n}\delta_{n}^{q}\bm{v}_{n} \\
\bm{F}_{f,t} = &-K_{t}\delta_{n}^{s}\bm{\delta}_t - \gamma_{t}\delta_{n}^{q}\bm{v}_{t} \
\end{align}
\end{subequations}
where $K_{n}$ and $K_{t}$ denote normal and tangential components of the stiffness parameter, and $\gamma_{n}$ and $\gamma_{t}$ denote normal and tangential components of the damping parameter. These parameters can be computed from material properties. The contact displacement vector is $\bm{\delta} = \bm{\delta}_{n} + \bm{\delta}_{t}$, and $\hat{\bm{n}} = \bm{\delta}_{n}/\|\bm{\delta}_{n}\|$ are the unit directions. Finally $\bm{v}_{n}$ and $\bm{v}_{t}$ denote the normal and tangential components of the velocity vector. For the Hertzian contact model used in this work, $p = 3/2$, $q=1/4$, and $s=1/2$. The tangential force respects the Coulomb friction condition $\|\bm{F}_{c}\| \leq \mu\|\bm{F}_{n}\|$ for friction coefficient $\mu$. The normal and tangential stiffness and damping coefficients are computed from material physical properties as below:
\begin{subequations}
\label{Params1}
\begin{align}
K_{n} = & \frac{4}{3}\overline{Y}\sqrt{\overline{R}}\\
K_{t} = & 8\overline{G}\sqrt{\overline{R}}\\
\gamma_{n} = & -2\sqrt{\frac{5}{6}}\beta\sqrt{\frac{3}{2}\overline{m}k_{n}}\\
\gamma_{t} = & -2\sqrt{\frac{5}{6}}\beta\sqrt{\overline{m}k_{t}}
\end{align}
\end{subequations}
where $\beta$ is computed from the coefficient of restitution $e \in [0,1]$:
\begin{equation}
\label{BetaFromE}
\beta = \frac{\ln{e}}{\sqrt{\ln^{2}{e} + \pi^{2}}}
\end{equation}
where $e\in[0,1]$ is the coefficient of restitution, with $\beta\to-1$ as $e\to0^+$ for a perfectly inelastic collision and $\beta=0$ at $e=1$ for a perfectly elastic collision.
The parameter $\overline{Y}$ is the effective Young's modulus and $\overline{G}$ is the effective shear modulus, $\overline{R}$ is the effective radius of curvature and $\overline{m}$ is the effective mass. These quantities are all computed for two bodies $i$ and $j$ in contact:
\begin{subequations}
\label{EffQuant}
\begin{align}
\overline{Y} = & \left( \frac{1 - \nu_i^2}{Y_i} + \frac{1 - \nu_j^2}{Y_j} \right)^{-1} \\
\overline{G} = & \left( \frac{2(2 + \nu_i)(1 - \nu_i)}{Y_i} + \frac{2(2 + \nu_j)(1 - \nu_j)}{Y_j} \right)^{-1} \\
\overline{R} = & \left(R_i^{-1} + R_j^{-1}\right)^{-1} \\
\overline{m} = & \left(m_i^{-1} + m_j^{-1}\right)^{-1}
\end{align}
\end{subequations}
where $Y_{i}$, $\nu_{i}$, $R_{i}$, $m_{i}$ denote Young's modulus, Poisson's ratio, radius of curvature, and mass of body $i$. We adopt a conservative integration time step of $\Delta t = 0.1\,t_{\mathrm{c}}$, where $t_{\mathrm{c}}$ is the Hertzian collision time \citep{Nesterenko2001}.

For further details of the contact detection and of mass element interaction modeling, the interested reader is directed to \cite{Fabio_Nbody}, and to the references of the underlying Chrono physics engine \citep{Chrono_conference2016,ChronoDEMPM}. GRAINS also has the ability to model cohesion, although cohesion forces are set to zero in this work. While in general the cohesiveness of rubble piles is still an open question, we have reason to expect that it can be near-zero, at least towards the surface of the body \citep{NoCohesion_Walsh22} wherein the bulk of dissipation is expected \citep{NimmoMatsuyama}.  

\subsubsection{N-body Dynamical Modeling}
We study the orbital and attitude dynamics of the rubble pile secondary in a binary asteroid system. In our numerical implementation, for timespans of interest of days to weeks, we are still limited to a few thousand particles in an $N$-body simulation. In lieu of splitting the budget of particles between primary and secondary, we instead opt for rendering the secondary as a rubble pile with the maximum practical resolution, and hence the primary is modeled as a single massive particle. This is appropriate for our studies which are focused specifically on dissipation within the secondary. A rough but useful perspective, assuming $M\gg m$, is that changes to the orbit of the secondary are dominated by dissipation within the primary, whereas changes in the spin state of the secondary are dominated by dissipation within the secondary \citep{SSD_1999}. 

The dynamics at work are the result of forces of $N$-body gravitation, inter-element contact, and friction. The $N$-body potential is given below for primary mass $M$ and mass elements $\text{d}m_{i}$, $i \in [1,N]$, neglecting the second-order terms from attraction between individual non-spherical elements:
\begin{equation}
\label{Nbody1}
U = -G\left( \sum_{j}\frac{M\text{d}m_{j}}{\delta r_{j}} + \sum_{1\leq i < j \leq N}\frac{\text{d}m_{i}\text{d}m_{j}}{\delta r_{ij}}\right)
\end{equation}
where $m = \sum_{i}\text{d}m_{i}$ is the total mass of the secondary and $G$ is the universal gravitational constant. By default, GRAINS does not compute the gravitational effects of non-spherical moments between mass elements, to avoid prohibitive runtimes. For consistency these effects should also be neglected in any computations of the potential. The neglected inter-element gravitational torque effects are quite sub-dominant in comparison to contact and friction effects, however, where the non-sphericity of individual particles is very important. Further details of the dynamics model can be found in \cite{Ferrari_2020}. The aggregate inertia tensor of the rubble pile secondary is computed without approximation, with the positions, angular velocities, and angular momenta of all individual particles also tabulated. As a result, the total angular momentum of the system is computed and tabulated. We observe that it is approximately conserved in our simulations to a high degree of accuracy (to be discussed -- see Results), establishing the first of two tests of GRAINS physical accuracy for the binary system.

Let us provide some final comments on the non-gravitational forces. The contact modeling was already discussed previously in Section 2.2.3. Note that the dynamics in this work neglect the spin effects of YORP \citep{Holsapple_YORP}. For our highly disturbed spin states, closely separated primary and secondary, and relatively short simulation timescales, the tidal torques will dominate. We also neglect the BYORP effect, which is important for long-term orbital evolution, but initiates only after the tidal locking of a binary secondary \citep{Cuk_BYORP}.

Total system energy, as computed in GRAINS, may be computed simply as the sum of kinetic, potential, and contact energies across all particles:
\begin{align}
\label{TotalETrue}
E(t) = \tfrac{1}{2}M\|\bm{v}_M\|^{2} + \sum_{j}\left[\tfrac{1}{2}m_{j}\|\bm{v}_{j}\|^{2} + \tfrac{1}{2}\bm{\omega}_{j}^{\top}\bm{I}_{j}\bm{\omega}_{j}\right]
 - G\left(\sum_{j}\frac{M m_{j}}{r_{j}} + \sum_{1\leq i<j\leq N}\frac{m_{i}m_{j}}{r_{ij}}\right) \nonumber \\
 + \sum_{k\in\mathcal{K}(t)} E_{k,\mathrm{strain}}
 + \sum_{k\in\mathcal{K}_{\mathrm{all}}(t)}\left(W_{k,\mathrm{damp}} + W_{k,\mathrm{slip}} - W_{k,\mathrm{coup}}\right)
\end{align}
where body $j$ has mass $m_{j}$, translational velocity $\bm{v}_{j}$, inertia tensor $I_{j}$, and angular velocity $\bm{\omega}_{j}$, $M$ is the mass of the central body, and $r_{j}$ and $r_{ij}$ are centre-to-centre distances. Again the non-spherical gravitational moment contributions are not tracked. The contact terms distinguish two sets. $\mathcal{K}(t)$ is the set of contacts active at time $t$, which store elastic energy. $\mathcal{K}_{\mathrm{all}}(t)$ is the set of all contacts formed up to time $t$, including those that have since broken, whose dissipated energy remains part of the budget. The expressions for the Hertzian-Coulomb model used in GRAINS are given by \citep{Direnzo2004, Sun2022}:

\begin{align}
E_{k,\mathrm{strain}} &= \tfrac{2}{5}K_{n,k}\,\delta_{n,k}^{5/2} + \tfrac{1}{2}K_{t,k}\,\delta_{n,k}^{1/2}\|\bm{\delta}_{t,k}\|^{2},\\
W_{k,\mathrm{damp}} &= \int_{t_k^{0}}^{t}\gamma_{n,k}\,\delta_{n,k}^{1/4}\|\bm{v}_{n,k}\|^{2}\,dt' + \int_{t_k^{0}}^{t}\gamma_{t,k}\,\delta_{n,k}^{1/4}\|\bm{v}_{t,k}\|^{2}\,dt',\\
W_{k,\mathrm{slip}} &= \int_{t_k^{0}}^{t}K_{t,k}\,\delta_{n,k}^{1/2}\,\bm{\delta}_{t,k}\cdot\left(\bm{v}_{t,k}-\dot{\bm{\delta}}_{t,k}\right)dt',\\
W_{k,\mathrm{coup}} &= \int_{t_k^{0}}^{t}\tfrac{1}{4}K_{t,k}\,\delta_{n,k}^{-1/2}\|\bm{\delta}_{t,k}\|^{2}\,\dot{\delta}_{n,k}\,dt',
\end{align}

where $t_k^{0}$ is the time at which contact $k$ formed, and the integrals run to $t$ or to the time at which the contact broke, whichever occurs first. Here $\bm{v}_{n,k}$ and $\bm{v}_{t,k}$ are the normal and tangential components of the relative velocity at the contact point, including the contribution of the angular velocities. $E_{k,\mathrm{strain}}$ is the elastic energy stored in the normal and tangential springs, while $W_{k,\mathrm{damp}}$ and $W_{k,\mathrm{slip}}$ are the energy dissipated by the normal and tangential dashpots and by Coulomb sliding, respectively. The relative tangential displacement $\bm{v}_{t,k}\,dt$ is split into an elastic part $d\bm{\delta}_{t,k}$, which is stored in the tangential spring, and a sliding part $\bm{v}_{t,k}\,dt - d\bm{\delta}_{t,k}$, whose work is lost. In the discrete implementation $\dot{\bm{\delta}}_{t,k}$ is discontinuous when the Coulomb condition is enforced, so $W_{k,\mathrm{slip}}$ is evaluated as the decrease in tangential strain energy when $\bm{\delta}_{t,k}$ is rescaled to satisfy $\|\bm{F}_{f,t}\| \leq \mu\|\bm{F}_{f,n}\|$. $W_{k,\mathrm{coup}}$ appears due to the specific tangential force law used here. Because the tangential stiffness $K_{t,k}\delta_{n,k}^{1/2}$ depends on the normal overlap, a change in $\delta_{n,k}$ while the tangential spring is loaded changes the stored tangential energy without any corresponding work by the contact forces, since the normal force does not depend on $\bm{\delta}_{t,k}$. The normal and tangential force laws are therefore not jointly conservative, and $W_{k,\mathrm{coup}}$ measures the energy injected or removed through this coupling. It enters Eq.~\eqref{TotalETrue} with a negative sign and vanishes for contacts with either no tangential load or a constant normal overlap. In our simulations we observe that the bulk of the energy lost from the secondary is transferred into the orbit, with some internal dissipation due to friction.

For a point-mass primary of mass $M$ and a perfectly rigid secondary of mass $m$ and principal moments of inertia $C\geq B \geq A$, the total system energy is given below:
\begin{equation}
\label{TotalEApp}
E_{\text{rigid}} = \frac{1}{2}Mv_{1}^{2} + \frac{1}{2}mv_{2}^{2} + \frac{1}{2}\bm{\omega}^{\top}I\bm{\omega} - \frac{GMm}{r} - \frac{GM}{2r^{3}}\left(A + B + C - 3\Phi(\bm{\eta},r)\right)
\end{equation}
where $I$ is the secondary inertia tensor, $\bm{\omega}$ is the secondary rotational angular velocity, $\bm{\eta}$ denotes the attitude of the secondary, and $\Phi$ is from a second-order expansion of the potential in MacCullagh's approximation \citep{SSD_1999}:
\begin{equation}
\label{Phi1}
\Phi(\bm{\eta},r) = \frac{Ax^{2} + By^{2} + Cz^{2}}{r^{2}}
\end{equation}
where the vector $-^{\mathcal{B}}\bm{r}=(x, y, z)^{\top}$ resolves the location of the perturbing primary in the body frame of the secondary, and $r = \sqrt{x^{2} + y^{2} + z^{2}}$. For tidally locked cases with limited deformations of the binary secondary, the moments of inertia do not change much\footnote{This convenient rigid-body approximation of total energy neglects small energy contributions of inter-particle movements and gravity. For our scenarios with relatively placid modifications in shape, they contribute negligibly to the total energy. For GRAINS scenarios with violent reshaping events or major disruptions, this approximation cannot be used.} and frictional dissipation is not aggressive, and this quantity thus changes fairly little over the course of a simulation. We can also isolate the \textit{rotational energy} of the secondary, which is a function of the orientation and rotational state and distance between the two bodies:
\begin{equation}
\label{Erot1}
\mathcal{E} = \frac{1}{2}\bm{\omega}_{r}^{\top}I\bm{\omega}_{r} + V(\bm{\eta}, r)
\end{equation}
where $\bm{\omega}_{r} = \bm{\omega} - \bm{\omega}_{\text{orb}}$ denotes the angular velocity relative to the rotating Hill frame, and $V(\bm{\eta},r)$ is the final term in Eq.~\eqref{TotalEApp}. Note in the planar case where rotations are limited to the orbital plane, it takes on the familiar form $V(\psi) = -\frac{3}{4}\frac{GM}{r^{3}}\left(B - A\right)\cos{2\psi}$ for yaw angle $\psi$ about the body polar axis \citep{SSD_1999,Goldreich_Peale_SpinOrb}. Applying the additional assumption of a circular Keplerian orbit $r = a = a_{0}$, neglecting the effect of attitude on the orbit and also the tidal dissipation, Eq.~\eqref{Erot1} becomes a conserved quantity. In reality, the orbit is not perfectly circular, and the orbital and attitude dynamics of the binary secondary are linked. Nonetheless, for our simulations with near-circular orbits, the equation is still useful because it allows us to obtain approximations of limits on the orientation and rotation state of the secondary, in particular whether roll is permitted, and at a high enough energy level, whether super-synchronous rotation (SSR) is expected. Consider first the attitude case $\psi = \pm\frac{\pi}{2}$, $\theta = \phi = 0$ with $\bm{\omega}_{r} = \bm{0}$. In this case, $\Phi(\bm{\eta}) = B$, the rotational kinetic energy term in Eq.~\eqref{Erot1} is zero, and the energy for this case is given below:
\begin{equation}
\label{ErotLim1}
\mathcal{E}_{\text{SSR}} = -\frac{GM}{2r^{3}}\left(A + C - 2B\right)
\end{equation}
Alternatively, for the attitude case $\psi = \theta = 0$, $\phi = \pm\frac{\pi}{2}$ with $\bm{\omega}_{r} = \bm{0}$, $\Phi(\bm{\eta}) = A$, yielding the following:
\begin{equation}
\label{ErotLim2}
\mathcal{E}_{\text{roll}} = -\frac{GM}{2r^{3}}\left(B + C - 2A\right)
\end{equation}
Starting with some initial orientation $\bm{\eta}(0)$ and spin $\bm{\omega}_{r}(0)$, the value of the resulting rotational energy $\mathcal{E}$ relative to the critical values $\mathcal{E}_{\text{SSR}}$ and $\mathcal{E}_{\text{roll}}$ provides limitations on the resulting qualitative behavior that is possible due to conservation of the rotational energy. Namely, roll is possible when $\mathcal{E} > \mathcal{E}_{\text{roll}}$, and not possible when $\mathcal{E} \leq \mathcal{E}_{\text{roll}}$. Similarly, super-synchronous rotation is possible when $\mathcal{E} > \mathcal{E}_{\text{SSR}}$, whereas tidal locking is expected when $\mathcal{E} \leq \mathcal{E}_{\text{SSR}}$. These conclusions rely on conservation of rotational energy and neglect orbit-attitude coupling, so they do not exactly hold in the simulation results. Nonetheless they are useful approximations that we make use of when choosing initial conditions to produce a certain type of behavior. By our convention $C\geq B \geq A$, it is also true that $\mathcal{E}_{\text{SSR}} \geq \mathcal{E}_{\text{roll}}$, which makes intuitive sense. For a good discussion of coupled orbital and attitude dynamics (assuming a rigid secondary and point mass primary), see \cite{Meyer_EnergyDissBinary}.

\subsubsection{Tidal Theory and Numerical Computations}
\label{subsec:theory}

To derive the tidal torque and energy dissipation, the Kaula formulation is used \citep{kaula1} with the additions of \cite{EfroimskyMakarov2013, Efroimsky2014} which ensure more correct frequency dependency in the tidal response, allowing this formulation to be used for a more general set of bodies and rheologies. The full detailed derivation of these quantities can be found in the aforementioned references, nevertheless the main points of these derivations are presented here for completeness.

A perturbing body $k$, of mass $M_{k}$ and located at $\bm r_{k}$, generates a disturbing potential at a point $\bm R$ on the surface of the perturbed body of radius $R<|\bm r_{k}|$:
\begin{equation}
W(\bm R,\bm r_{k}) \;=\; \sum_{l=2}^{\infty} W_l(\bm R,\bm r_{k})
\;=\; -\,\frac{GM_{k}}{r_{k}}\sum_{l=2}^{\infty}\left(\frac{R}{r_{k}}\right)^{l} P_l(\cos\gamma) ,
\label{eq:W1}
\end{equation}

where $P_l$ is the Legendre Polynomial and $\gamma$ the angle between $\bm r_{k}$ and $\bm R$, both measured from the perturbed body's center. Using spherical coordinates, with $r$ the radial distance, $\phi$ the latitude from the perturbed body's equator, and $\lambda$ the longitude measured from a fixed meridian, the potential can be written using the associated Legendre Polynomials $P_{lm}(\cos \phi)$ as:

\begin{equation}
W(\bm R,\bm r_{k}) = -\frac{GM_{k}}{r_{k}}\sum_{l=2}^{\infty}\left(\frac{R}{r_{k}}\right)^{l}
\sum_{m=0}^{l}\frac{(l-m)!}{(l+m)!}\,(2-\delta_{0m})\,P_{lm}(\sin\phi)\,P_{lm}(\sin\phi_{k})\,
\cos m(\lambda-\lambda_{k}) .
\label{eq:W2}
\end{equation}

\cite{kaula1} introduced a transformation of this potential into the Keplerian elements $\bm r_{k}=(a_{k},e_{k},i_{k},\Omega_{k},\omega_{k},\mathcal M_{k})$ of the perturber, while keeping the surface point $\bm R=(R,\phi,\lambda)$ fixed:

\begin{align}
W(\bm R,\bm r_{k}) &= \sum_{lmpq} W_{lmpq} \nonumber\\
&= -\frac{GM_{k}}{a_{k}}\sum_{l=2}^{\infty}\left(\frac{R}{a_{k}}\right)^{l}
\sum_{m=0}^{l}\frac{(l-m)!}{(l+m)!}\,(2-\delta_{0m})\,P_{lm}(\sin\phi)
\sum_{p=0}^{l} F_{lmp}(I) \sum_{q=-\infty}^{\infty} G_{lpq}(e_{k})\nonumber\\
&\qquad\times
\begin{Bmatrix}\cos\\ \sin\end{Bmatrix}_{l-m\ \text{odd}}^{l-m\ \text{even}} \Big(v_{lmpq,k} - m(\lambda+\theta^{*})\Big),
\label{eq:W3}
\end{align}

where $\theta^{*}$ represents the rotational angle of the perturbed body, and $F_{lmp}(I)$ and $G_{lpq}(e_{k})$ are the inclination functions and the eccentricity polynomials, respectively, which are expansions in terms of their respective orbital elements. Their full expression can be found in \cite{SSD_1999}. Here, instead of the inclination $i_k$ the relative obliquity $I$ is used in case the relative spin is not aligned with the orbital normal vector. The linear combination $v_{lmpq,k}$ is given by:

\begin{equation}
v_{lmpq,k} \equiv (l-2p)\,\omega_{k} + (l-2p+q)\,\mathcal M_{k} + m\,\Omega_{k}.
\label{eq:v}
\end{equation}

Linearising the argument of \eqref{eq:W3} over a timescale short compared with the apsidal-nodal period gives:

\begin{equation}
v_{lmpq,k} - m(\lambda + \theta^{*}) = \omega_{lmpq}(t-t_0) - m\lambda + v_{lmpq,k}(t_0) - m\theta^{*}(t_0),
\label{eq:linv}
\end{equation}

with the following representing the tidal Fourier modes:

\begin{equation}
\omega_{lmpq} \;\equiv\; \dot v_{lmpq,k} - m\dot\theta_{k}
\;=\; (l-2p)\,\dot\omega_{k} + (l-2p+q)\,\dot{\mathcal M}_{k} + m\big(\dot\Omega_{k}-\dot\theta_{k}\big).
\label{eq:mode}
\end{equation}

Here, $\dot{\mathcal M}_{k}$ is the (anomalistic) mean motion, where the following definition is used $n \equiv \dot{\mathcal M}$. The positive definite forcing frequency is also defined as follows:

\begin{equation}
    \chi_{lmpq} = |\omega_{lmpq}|
    \label{eq:chi}
\end{equation}

The perturbing potential distorts the perturbed body's shape; the resulting change to the perturbed body's own exterior potential $U(\bm z,\bm z_{k})$ at a point $\bm z$ is, in the elastic (frictionless) limit,

\begin{equation}
U(\bm z,\bm z_{k}) = \sum_{l=2}^{\infty} U_l(\bm z)
= \sum_{l=2}^{\infty} k_l\left(\frac{R}{z}\right)^{l+1} W_l(\bm R,\bm z_{k}),
\label{eq:U-static}
\end{equation}

where $k_l$ are the static Love numbers. The reaction of the perturbed body is not instantaneous due to internal friction and energy dissipation, which results in a lagging reaction to the tidal perturber. It is important to note, that this lagging depends on frequency \cite{EfroimskyMakarov2013} and thus each mode needs to have its own lag and love number. To include this, a dynamical love number $\hat{k}_l$ is substituted into Eq. \eqref{eq:U-static}. This is a linear operator that delays each Fourier mode of Eq. \eqref{eq:U-static} separately:

\begin{align}
U(\bm z,\bm z_{k}) &= -\frac{GM_{k}}{a}\sum_{l=2}^{\infty}\left(\frac{R}{z}\right)^{l+1}
\left(\frac{R}{a}\right)^{l}\sum_{m=0}^{l}\frac{(l-m)!}{(l+m)!}(2-\delta_{0m})\,P_{lm}(\sin\phi)
\sum_{p=0}^{l}F_{lmp}(I)\sum_{q=-\infty}^{\infty}G_{lpq}(e)\nonumber\\
&\qquad\times k_l(\omega_{lmpq})
\begin{Bmatrix}\cos\\ \sin\end{Bmatrix}_{l-m\ \text{odd}}^{l-m\ \text{even}}
\Big(v_{lmpq} - m(\lambda+\theta) - \epsilon_l(\omega_{lmpq})\Big),
\label{eq:U-lagged}
\end{align}

where $k_l(\omega_{lmpq})$ and $\epsilon_l(\omega_{lmpq})$ are the frequency dependent tidal love number and phase lag\footnote{For near-spherical bodies both $k_l$ and $\epsilon_l$ are only functions of the $l$ mode and the dependency with respect to the other modes $m$, $p$, and $q$ only comes from the functional dependency to $\omega_{lmpq}$. When looking at triaxial ellipsoids, it was shown that the functional dependency of $k$ and $\epsilon$ also includes $m$, $p$, and $q$. In this work, it will be numerically shown that one mode dominates over all others and the analysis will consider this. Hence, this nuance is not included for ease-of-notation.}. The $k$ subscript for the orbital elements is dropped from here on out for notational simplicity. 

In most of the literature, lagging is parametrized through a tidal quality factor $Q$ rather than directly through the phase lag. It is defined through the energy lost per cycle relative to the peak energy stored in the distortion,

\begin{equation}
Q(\chi) = \frac{2\pi \mathcal{E}_{0}(\chi)}{\Delta \mathcal{E}(\chi)},
\label{eq:Q-def}
\end{equation}

which is linked to the phase lag by

\begin{equation}
\frac{1}{Q_l} \equiv |\sin\epsilon_l|,
\end{equation}

such that

\begin{equation}
\frac{k_l}{Q_l} = k_l(\chi)\,\sin\epsilon_l(\chi) .
\label{eq:kQ}
\end{equation}

Now that a formal expression of the tidal potential and its associated quantities is given, the tidal torque and energy dissipation can be found. The polar torque (i.e.\ the component orthogonal to the perturbed body's equator) can be obtained as the derivative of the tidal potential with respect to the sidereal angle\footnote{The torque due to $U$ on an exterior body is $T_z = -M_k\frac{\partial U}{\partial \theta_k}$, which creates a back-reaction on the tidally perturbed body of $\Gamma_{z} = -T_z$.}:

\begin{equation}
\Gamma_{z, tidal} = M_k\,\frac{\partial U}{\partial \theta}.
\label{eq:torque-def}
\end{equation}

Averaging over one cycle, one can find (assuming no precession of the node $\dot{\Omega}_k\approx 0$) \citep{Renaud2021}:

\begin{align}
\langle \Gamma_{z,tidal} \rangle
&= \frac{GM_k}{a}\sum_{l=2}^\infty\left(\frac{R}{a}\right)^{2l+1}
\sum_{m=0}^{l}\frac{(l-m)!}{(l+m)!}(2-\delta_{0m})\sum_{p=0}^{l}F_{lmp}^2(I)\sum_{q=-\infty}^{\infty}G_{lpq}^2(e) \nonumber\\
&\qquad  \times  m\,\Sgn(\omega_{lmpq})\,k_l(\omega_{lmpq})/Q_l(\omega_{lmpq}).
\label{eq:torque_total}
\end{align}

The total torque acting on the body can be split into two components: the torque due to \textit{permanent mass asymmetry} $\bm{\Gamma}_{\text{PMA}}$, and the torque due to a time-varying effect on the MoI induced by the gravitational pull of the primary, $\Gamma_{\text{Tidal}}$, given in Eq. \eqref{eq:torque_total}. On terrestrial planets, the former term is due to the gravitational torque on \textit{static} large-scale inhomogeneities and uncompensated topographic variations \citep{Melosh,SSD_1999}, and the latter is due to the torque on the \textit{dynamic} lagging transient deformation of the body from the gravitational pull of a perturber. The relative strength of these two torques determines the final spin state of the body \citep{Goldreich_Peale_SpinOrb}. In the case of rubble pile asteroids, the body has very low effective ``strength" or rigidity, held together by weak gravitational, friction, and perhaps electrostatic forces \citep{Sanchez2014}. For such low-strength granular bodies, this concept may be harder to visualize, but it still holds. 

In this paper we are interested in how dissipative the rotational settling process is, by which the binary secondary achieves an equilibrium spin state from an initially disturbed state. We will explore both planar spin and tumbling cases for our initially unsettled spin states. Consider the fully planar case of supersynchronous rotation, for which $\psi$ is unbounded. In this case, for a simple circular orbit, the rotationally averaged torque due to the permanent mass asymmetry is zero. Thus the rotationally averaged total torque is just a function of the tidal torque:

\begin{equation}
\label{TidalTorque3}
\langle \Gamma_{z}\rangle = \frac{1}{2\pi}\int_{-\pi}^{\ \pi}\left(\Gamma_{z,\text{PMA}}(\psi) + \Gamma_{z,\text{tidal}}(\psi,t)\right)\text{d}\psi = \langle \Gamma_{z,\text{tidal}} \rangle
\end{equation}

From the GRAINS results, we compute and tabulate the total torque $\bm{\Gamma}(t)$ at various times. In the case of planar spin, numerically averaging e.g. by a trapezoidal or spline method, we are also able to estimate the averaged torque, Eq.~\eqref{TidalTorque3}. In this work we directly numerically compute the tidal torque on the secondary, due to its own internal dissipation, and are not beholden to any particular tidal theory. It is however of great interest to represent the results we recover by well-known metrics in the literature. 

For tumbling and non-principal axis rotation cases, two assumptions break. First, $\dot{\psi}$ is not constant anymore and thus the PMA torque does not cancel out anymore. Second, a full 2$\pi$ rotation just in $\psi$ does not necessarily constitute a return to the original orientation anymore. We seek to directly compare the dissipation in disrupted non-planar spin states to the super-synchronous spin case, and thus need to adjust the approach for this case. This is achieved by leveraging Newton's equation for rotational motion $\dot{\bm{H}} = \bm{\Gamma}$ and averaging over time as follows:

\begin{equation}
\label{TidalTorque4}
\langle \Gamma_{z}\rangle = \lim_{T* \to \infty}\frac{1}{T^*}\int_{t_0}^{\ t_0 + T^*}\left(\Gamma_{z,\text{PMA}}(t) + \Gamma_{z,\text{tidal}}(t)\right)\text{d}t = \langle \Gamma_{z,\text{tidal}} \rangle.
\end{equation}

The main idea behind this approach is that $\Gamma_{z,\text{PMA}}$ is non-dissipative and continuously exchanges momentum between the secondary's spin and the orbit \cite{SSD_1999}. Hence, this torque remains bounded over time, whereas $\Gamma_{z,\text{tidal}}$ secularly changes. Thus, if an appropriate, and appropriately large $T^{*}$ is selected, the $\Gamma_{z,\text{PMA}}$ will be averaged out.

The other observable discussed here is the energy lost due to internal friction caused by tidal forces. An analytical expression for the tidal dissipation of a body that is in a synchronous rotation where only the first-order terms are included are well known and can be found in e.g. \cite{Peale:1999} and \cite{SSD_1999}. However, for a more general expression of bodies that can be in any rotational state, the derivation used in \cite{Efroimsky2014, Frouard2017, Efroimsky2018} needs to be used. Including the full derivation is out of the scope of this paper and interested readers are directed to the previously mentioned references. The expression in terms of the tidal potentials can be found to be as follows:

\begin{equation}
\langle \dot E\rangle = \frac{1}{4\pi GR}\sum_{l=2}^{\infty}(2l+1)\int \big\langle W_l\,\dot U_l\big\rangle\, \dd S,
\label{eq:P-surface}
\end{equation}

with $\dd S$ an element of the undeformed surface of the body. It can then be shown that the tidal dissipation rate (away from any spin-orbit resonance) is given by:

\begin{equation}
\langle \dot E\rangle = \frac{GM_k^{2}}{a}\sum_{l=2}^\infty\left(\frac{R}{a}\right)^{2l+1}
\sum_{m=0}^l\frac{(l-m)!}{(l+m)!}(2-\delta_{0m})\sum_{p=0}^l F_{lmp}^2(I)\sum_{q=-\infty}^\infty G_{lpq}^2(e)\,
\chi_{lmpq}\,k_l(\chi_{lmpq})\sin\epsilon_l(\chi_{lmpq}).
\label{eq:P-spectral}
\end{equation}

Putting the torque \eqref{eq:torque-def} and the dissipation rate \eqref{eq:P-spectral} into a single matrix expression gives:
\begin{align}
\begin{bmatrix}\langle \Gamma_z\rangle\\ \langle \dot E\rangle\end{bmatrix}
&= \frac{GM_k}{a}\sum_{l=2}^\infty\left(\frac{R_j}{a}\right)^{2l+1}
\sum_{m=0}^{l}\frac{(l-m)!}{(l+m)!}(2-\delta_{0m})\sum_{p=0}^{l}F_{lmp}^2(I_j)\sum_{q=-\infty}^{\infty}G_{lpq}^2(e)\nonumber\\
&\qquad\times
\begin{bmatrix} m\,\Sgn(\omega_{lmpq})\,k_l/Q_l \\[2pt] \chi_{lmpq}\,M_k\,k_l/Q_l\end{bmatrix} .
\label{eq:matrix-general}
\end{align}
 
Restricting to the quadrupole ($l=2$) term only, and collapsing the $(m,p,q)$ triple sum into a single sum over a tidal-mode index $i$:

\begin{equation}
\begin{bmatrix}\langle \Gamma_z\rangle\\ \langle \dot E\rangle\end{bmatrix}
= \frac{GM_k^2 R^5}{a^6}\sum_{i}\frac{(2-m)!}{(2+m)!}(2-\delta_{0m})\,F_i^2\,G_i^2
\begin{bmatrix} m\,\Sgn(\omega_i)\,k_2/Q\\[2pt] \chi_i\,k_2/Q\end{bmatrix}.
\label{eq:matrix-l2}
\end{equation}

By using the appropriate averaging as discussed before and inverting the equations, the $k_2/Q$ (where the 2 subscript on Q is dropped to relate this with the wider literature) can be extracted using either averaged torque or average energy rate as observables. Considering just the first tidal mode, for instance, an estimate for $Q/k_{2}$ can be isolated as
\begin{equation}
\label{Qk2_new}
\frac{Q}{k_2} \sim -\left(\frac{2}{3}\frac{a^6}{GM^2}R^{-5}\langle\Gamma_z\rangle\right)^{-1}.
\end{equation}

\subsection{Capabilities and Limitations}
\subsubsection{Numerical modeling in GRAINS}
Runtime timescales are long in $N$-body nonspherical granular mechanics-based simulations due to the computationally costly contact-checking operations \citep{Ferrari_2020,Fabio_Nbody}. However, the simulation of contact between non-spherical elements allows us to simulate certain granular mechanisms resulting from non-sphericity of the constituent particles (see e.g. \cite{Fabio_IcarusTop} and references therein). In our current implementation of GRAINS, we can simulate models with thousands of particles for $\sim$50 orbits for a Dimorphos-like orbit period of $\sim$12 hrs, and beyond this is not practical for studies with reasonable turn-around times. 
However, this is enough for meaningful exploration. Using the rotational energy equation Eq.~\eqref{Erot1} to specify reasonable starting states (say, not far above $\mathcal{E}_{\text{SSR}}$ for the super-synchronous case) enables interesting qualitative behavioral transitions (such as spin-down) to be set up and observed. 

\begin{figure}[h!]
\centering
\includegraphics[scale=0.8]{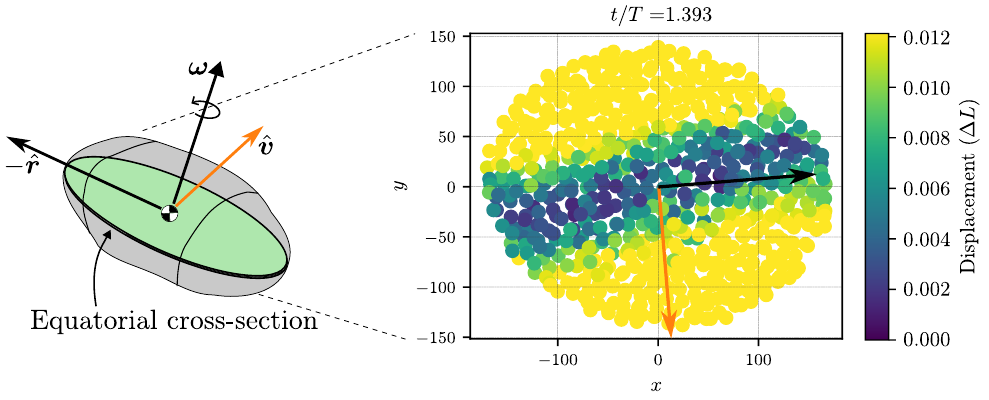}
\caption{Modeling of each discrete element of the rubble pile enables advanced study of local and large-scale rock movements. Here, an equatorial cross-section ``snapshot" displays the color-coded motion of individual grains, clearly illustrating the resultant degree-2 tidal distortion nearly aligned with the radial vector, with misalignment due to tidal lag. Note that each mass element is represented as a colored circle -- our discrete element model does not use spherical mass elements.}
\label{fig:asteroidCS1}
\end{figure}

The tracking of individual mass elements enables unique local and global studies of rock movements -- settling, jostling, migration, and global tidal response. This is illustrated conceptually by Fig.~\ref{fig:asteroidCS1} which depicts, for our Dimorphos-like body, the displacement of rocks in an equatorial cross-section due to tidal response in a super-synchronous spin simulation. Note that GRAINS makes use of a non-dimensional length scale (units of $L$) as a consequence of the methodology for accommodating specified target bulk densities. The interested reader is referred to \cite{Fabio_Nbody}. The key numerical results in this paper are presented to SI units. 

While limited by the tradeoff between number of mass elements and simulated timespan, GRAINS is particularly well-suited to modeling rubble pile secondaries such as Dimorphos. The reasons are twofold. First, because modeling the small secondaries allows for a finer resolution (e.g. smaller boulder size) for the same number of particles in comparison to the much larger primary. Assuming a body constructed of similarly sized mass elements, individual mass elements scale in size as $R_{\text{elem}} \sim \frac{R}{N^{1/3}}$ for secondary mean radius $R$. Thus for a Dimorphos-like model with a few thousand elements, the resulting elements are meter-scale. Second, observed disparities in cohesive strength between Didymos and Dimorphos suggest that fine grains may be scarce in binary secondaries \citep{Dimoprhos_from_DART}, possibly due to fine grain escape from solar radiation pressure before its accretion \citep{Dimoprhos_from_DART,Ferrari_PSJ2022}. \cite{DalyEtAlNature} also notes a steep drop-off in Dimorphos' surface boulder population for boulders with $<0.3\text{ m}$ width, noting the finding is not an artifact of limits on camera resolution.

\subsubsection{Emergence of Semidiurnal Tides}
\begin{figure}[h!]
\centering
\includegraphics[scale=1.1]{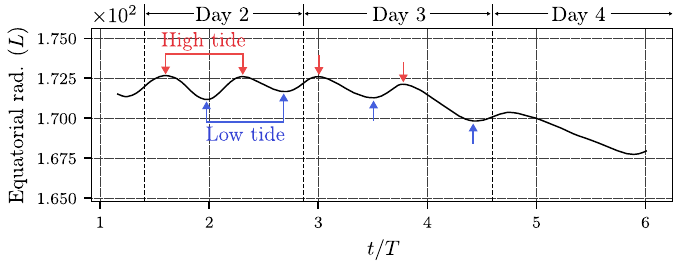}
\caption{Radial distance of a mass element from the barycenter computed over time, longitude $\varphi = 168^{\circ}$. Local maxima are marked with red arrows, and local minima with blue. In the final rotation, where the spin rate decreases significantly, there is significant equatorial radial settling.}
\label{fig:sdt0}
\end{figure}

We may directly observe the semidiurnal tide effects during supersynchronous rotation of a modeled rubble pile secondary. 
To illustrate this and to motivate the numerical studies of Section 3, consider a test case similar to the one used for Figure~\ref{fig:asteroidCS1}, initiated close to tidal locking to illustrate the disruptive effects of the tidal locking process. At a given topographic longitude on the secondary, twice per ``day" (e.g. full rotation in $\psi$), there will be two high and low tides acting on the regolith, corresponding to times when the effect of the tides from the primary are at their local maximum and minimum. 
Switching to our topographic frame $\mathcal{G}$, we resolve the radial distance from the barycenter of a mass element on the surface along the equator. The result is shown in Fig.~\ref{fig:sdt0}, which clearly depicts two local maxima and two local minima per day on the binary secondary. As the body spins down and approaches tidal locking, the global settling overwhelms this tidal effect, hence the qualitatively distinct behavior on Day 4. 

\begin{figure}[h!]
\centering
\includegraphics[scale=1.1]{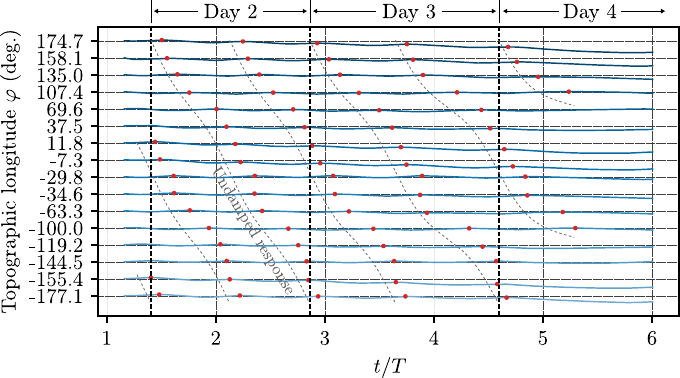} 
\caption{Timing of semidiurnal high tide vs. longitude. For select longitudes along the equator, the radial distance of a mass element from the barycenter is computed over time. Local maxima are marked with red dots, revealing a global order-2 pattern. The gray lines approximate mean timings of high tide for a perfectly elastic body whose response to the tidal forcing is instantaneous. Tidal temporal lag can be visualized via the horizontal separation of high tide points from the nearby diagonal lines. Note the variations with longitude $\varphi$ and time (or relatedly, spin rate $\omega_{2,z}$).}
\label{fig:sdt1}
\end{figure}

We can confirm that the radial range variations of mass elements along the equator are indeed the semidiurnal tide by producing similar plots to Fig.~\ref{fig:sdt0} for a number of mass elements at different longitudes along the equator, recording the timing of the local maxima, and stacking these variations by longitude. The result is given in Fig.~\ref{fig:sdt1}. This figure shows that the timing of high tide varies predictably as a function of longitude (similar to how the timing of the lunar semidiurnal tide varies based on longitude on the Earth), revealing the global extent of the tide. From the horizontal distance between high tide markers and the horizontal dashed lines (representative of the response of a completely non-dissipative body), the temporal lag of the semidiurnal tide can be inferred. Note that it is not a constant function of longitude. Furthermore, comparing the temporal lags across Days 2 and 3, we can discern that this lag is also a function of spin rate $\omega_{2,z}$.  
Lastly, the tidal strain can be estimated, in this illustration in the range $10^{-4} \lesssim \frac{H}{R} \lesssim 10^{-3}$ for radial variations $H$ and local equatorial radius $R$.

\subsubsection{Tidal dissipativity}
We assume that for small rubble pile secondaries such as Dimorphos, tidal dissipation is dominated by friction, in concurrence with \cite{NimmoMatsuyama}, as opposed to plastic yielding of stress concentrations due to roughness of the mass elements (e.g. \cite{GoldreichSari}). Our reasoning is that the secondaries are small and loosely packed, so shallow mass elements can shift and rotate subject to local stresses. For mass elements not in a maximally compact configuration, contacts on stress concentrations could exert a torque on the element that overcomes the resistive torque of contact friction with neighboring elements. Thus reconfiguration and compaction can occur before plastic yielding, and this process removes energy via frictional sliding. 

We're interested in estimating the dissipativity of rubble pile binary secondaries. Cognizant of the limitations of numerical modeling, we prefer an underestimate to an overestimate. 
Real rubble pile binary secondaries, while possibly deficient in fine regolith, will surely have more fine regolith than our simulation with meter-scale mass elements. Classically one would expect that they should be more dissipative than our model. Colloquially, a bag of marbles is more dissipative than a single large marble of the same material and mass \citep{NimmoMatsuyama}. Similarly, a bag of ``sand" of the same material and mass might be expected to be the most dissipative of the three. However, the relationship might not be this simple. In mixed regolith with low mass fractions $\mathcal{F}$ of small particles, where $\mathcal{F} = m_{s}/(m_{b} + m_{s})$ for small particles of total mass $m_{s}$ and large particles of total mass $m_{b}$, the rigidity/shear modulus increases with $\mathcal{F}$ until a critical point, as voids between the large particles are filled \citep{Murdoch_preprint2024}. This should have an overall effect of stiffening the rubble pile, reducing the thickness of the dissipative regolith layer. As the small particles begin to dominate, however, the fine regolith fraction becomes increasingly responsible for the total tidal dissipation. Thus the dissipative layer depth may decrease with finer grains, but its dissipativeness per unit volume may increase. In reality the expected dissipativeness of a rubble pile asteroid constructed from a given regolith is a complicated issue necessitating further experiments. 

In this work, we explore the case of a perturbed secondary. For this case, some of the scaling laws in \cite{NimmoMatsuyama} need to be re-derived, because they were originally derived for the case of studying dissipation in a primary by an orbiting secondary of smaller mass. We don't include these because the modifications of the needed equations are fairly easy. 
Second, much of the existing tidal theory rests on the assumption that, in bulk, such bodies behave in a way that can be approximated by viscoelastic theory. However, it is more likely to say that the bulk behavior will be approximated well, within certain dynamical regimes, with linearizable (though \textit{not necessarily classically viscoelastic}) equations with effective rigidity and viscosity, and possibly other unknown parameters \citep{Efroimsky_binaries}. Additionally, there is little work into the threshold of how many bodies (i.e. how many $N$ in an $N$-body simulation) are needed to render linearizable equations well-approximated by continuum mechanics, nor the effect of the scale of $N$ on the resulting continuum approximation. 

\section{Results}
\label{sec:results}
We provide the key results of simulations of the attitude, orbit, and rock movements of rubble pile secondaries in non-settled attitude states. The first is a case of super-synchronous rotation (denoted ``SSR") in which the secondary undergoes noticeable spin-down until tidal locking in the course of the simulation, and the second is a case of non-principal axis rotation (denoted ``NPA"). 
\subsection{Case 1: Super-synchronous rotation}
Important simulation information is summarized in Table~\ref{table:caseInfo}. 
The justification of free parameters in Table~\ref{table:caseInfo} is as follows. Quantities $M$ and $m$ are chosen to render a system similar to Didymos-Dimorphos. Material density is roughly what is expected for Dimorphos based on estimated bulk density and porosity ranges \citep{Dimoprhos_from_DART}. 

\begin{table}[h!]
\centering
\caption{Simulation parameters for the different simulations.}
\label{table:caseInfo}
\begin{tabular}{ll}
\hline
Parameter & Value \\ \hline
Mass Primary & $5.4\times 10^{11}$ kg \\
Secondary extents & (172.5, 168.6, 116.0) m \\
Bulk Density Secondary & $2400$ kg/m$^3$ \\
Material Density Secondary & $4453.387$ kg/m$^3$ \\
Secondary no. particles & 1982 \\
Particle Size-frequency Distribution & 5-20 m, power law with $-3$ exponent \\
Physical properties & Friction $\mu = 0.6$, Poisson $\nu = 0.3$, Restitution $e=0.5$ \\
Young's modulus & $Y = 5.0$ MPa \\
Initial attitude of secondary & $\psi = \theta = \phi = 0$ \\
Initial angular velocity & Case 1: $\bm{\omega}_{2} = (0,\, 0,\, 2n_{0})$ rad/s \\
 & Case 2: $\bm{\omega}_{2} = (9\times 10^{-5},\, 0,\, 1.5075\times 10^{-4})$ rad/s \\
Initial orbit & $a = 1.165$ km, $n_{0} = 1.456 \times 10^{-4}$ rad/s, $e = 0.025$, $f = 0^{\circ}$ \\
Orbital period & $\text{T}_{\text{orbit}} \approx 12.0$ hrs \\ \hline
\end{tabular}
\end{table}

The number of particles is $\sim$2K, and the average particle diameter is $\sim$10.3m. The initial attitude is set to zero for simplicity, and the angular velocity is chosen to render a super-synchronous spin state. The number of particles is chosen as a balance between fidelity and computational timespan.\footnote{For a model with $\sim$3K particles of similar size, the average width of the constituent particles of $\sim$10.3m is larger than the real boulder size limit, which drops off at around 0.3m \citep{DalyEtAlNature}, but a simulation with $\sim$1m sized boulders (a factor 10 reduction in size) would require $\sim10^{3}\times$ more particles (due to the scaling $R_{\text{elem}} \sim R_{\text{body}}/N^{1/3}$), with proportional increase in runtime.} 
The physical properties are chosen based on expected values of rock where possible: $\mu \sim 0.6$, $0.2 < \nu \leq 0.4$ \citep{PoissonRocks, SundayValidateDEM}. 

The Young's modulus is set below expected values for asteroid granular material, as a compromise between physicality and numerical feasibility of the contact simulations (i.e. the highest value allowing simulation runtimes $\sim$1 wk or less). Although exact properties are unknown, see discussion in \cite{Mohlmann2018} for porous regolith of comets, or \cite{Micromechanical_Rock} for values from laboratory experiments with various types of rock. As the Young's modulus of a simulated material is increased, the duration of collisions decreases, and the necessary numerical integration time step $\text{d}t$ to capture this effect accurately scales as $\text{dt} \propto Y^{-0.4}$ \citep{Tancredi_GranularDEM}. We adopt a reduced Young's modulus of $Y=5$ MPa to increase contact durations and permit larger integration time steps, consistent with previous DEM studies \citep{DeMartini2019,DellaGiustina2024,Fodde2026}. Previous tests have shown that this reduced Young modulus should not have effects on granular flow rates and measured angle of repose, but could have effect on bulk stiffness \cite{LOMMEN2014107}. 

\begin{figure}[h!]
\centering
\begin{subfigure}[b]{0.45\textwidth}
\centering
\includegraphics[width=3.1in]{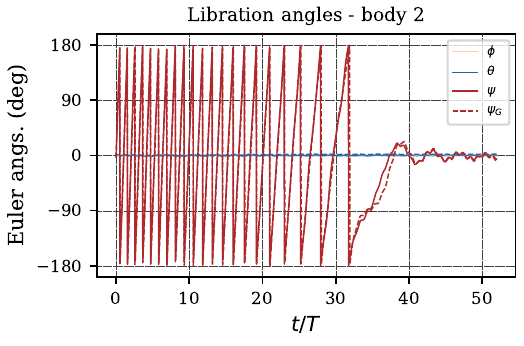} 
\caption{Euler angles of the principal axis body frame $\mathcal{B}$ (solid) and topography frame $\mathcal{G}$ (dashed).}
\label{fig:subfigB1}
\end{subfigure} 
\hspace{0.27in}
\begin{subfigure}[b]{0.45\textwidth}
\centering
\includegraphics[width=3.1in]{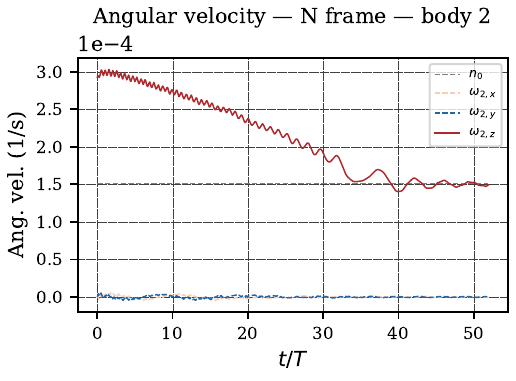} 
\caption{Angular velocities of the body frame $\mathcal{B}$, with initial mean motion indicated by $n_{0}$.}
\label{fig:subfigB2}
\end{subfigure}
\vspace{0.2in} 
\begin{subfigure}[b]{0.45\textwidth}
\centering
\includegraphics[width=3.1in]{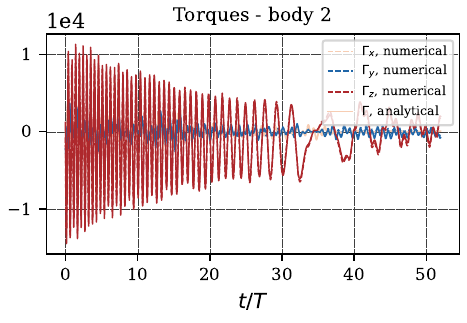} 
\caption{Components of tidal torques on the rubble pile secondary. Numerically recovered (dashed) and analytical approximation (solid).}
\label{fig:subfigB3}
\end{subfigure} 
\hspace{0.27in}
\begin{subfigure}[b]{0.45\textwidth}
\centering
\includegraphics[width=3.1in]{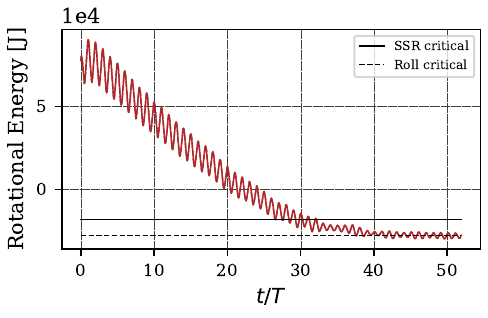} 
\caption{Rotational energy of the secondary, as computed by Eq.~\eqref{Erot1}, for the case of super-synchronous rotation.}
\label{fig:subfigB4}
\end{subfigure}
\caption{Simulation results for Case 1 (SSR). $\mathcal{B}$ frame Euler angles (a) correspond to the rotation matrix $[\mathcal{BO}']$, whereas $\mathcal{G}$ corresponds to the rotation matrix $[\mathcal{GO}']$. The angles indicate a primarily flat spin, coplanar with the orbit. The tidal torques (c) are provided in the $\mathcal{B}$ frame. Rotational energy (d) starts in a super-synchronous rotation case, as evidenced by $\mathcal{E} > \mathcal{E}_{\text{SSR}}$, but transitions to tidal locking at around $t/T\sim 30$.}
\label{fig:four_figures2}
\end{figure}

The timescales are computed from orbital parameters and the moments of inertia of the rubble pile model. The rubble pile secondary is initialized in a super-synchronous spin state, with the initial yaw rate (at the ``zero" attitude state with $\psi = \theta = \phi = 0$) equal to twice the initial mean motion, and the yaw angle is unbounded. Over the course of the simulation, the other two angles remain nearly zero so it is appropriate to consider this problem as 2D in our subsequent tidal analysis. The attitude evolution is given in terms of Euler angles in Fig.~\ref{fig:subfigB1} and in terms of the angular velocity of the $\mathcal{B}$ frame with respect to $\mathcal{O}'$ in Fig.~\ref{fig:subfigB2}. Note the secular decrease of the $z$ component of the angular velocity of the secondary, and the change in characteristic behavior.

Fig.~\ref{fig:subfigB3} gives the tidal torque about the $x,y,z$ body axes of the binary secondary, as reproduced from the outputs of GRAINS. Because the secondary spin remains nearly coplanar with the orbit, the torque is dominated by the polar ($z$) term. 
As a sanity-check, strong agreement can be noted between the numerically recovered estimate of the tidal torque (dashed lines), and the computation based on MacCullagh's approximation (solid lines). The polar torque shown is a sum of the conservative tidal torque $\Gamma_{z,\text{PMA}}$ and the dissipative tidal torque $\Gamma_{z,\text{Tidal}}$, and averaging over full rotations reveals a net negative (de-spinning) torque per rotation. This is to be used shortly for calculating the dissipativeness of the rubble pile. Fig.~\ref{fig:subfigB4} gives the rotational energy of the secondary, as computed by Eq.~\ref{Erot1}, for the case of super-synchronous rotation. This energy quantity displays significant secular decrease over the course of the simulation, and tidal locking occurs at around $t/T\sim 30$. 

\begin{figure}[h!]
\centering
\includegraphics[]{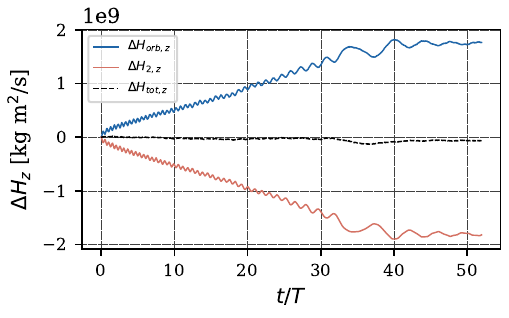}
\caption{The total angular momentum change over the whole simulation duration. The change in total angular momentum with respect to the initial value is at maximum around 0.015\%, which occurs just after the transition from SSR to a spin-locked state.}
\label{fig:Hconservation}
\end{figure}

\begin{figure}[h!]
\centering
\includegraphics[]{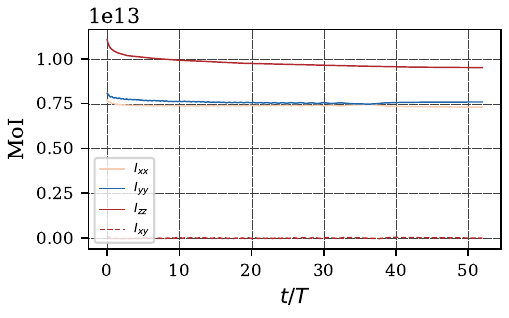}
\caption{Select moments of inertia, $\mathcal{G}$ frame. The very small but nonzero $I_{xy}$ demonstrates that it is not a principal axis frame.
Note the transient shape effects in the first orbits, and the longer-term settling of $I_{zz}$ and $I_{yy}$ due to spin-down.}
\label{fig:MoIs2}
\end{figure}

\begin{figure}[h!]
\centering
\includegraphics[]{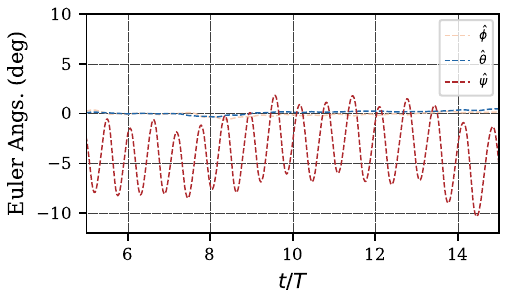}
\caption{Rotation, in 3-2-1 Euler angles, of $\mathcal{G}$ frame from an inertia tensor diagonalizing principal axis frame. These angles are computed from the data used to make Fig.~\ref{fig:MoIs2}. Note the clear semidiurnal signal in $\hat{\psi}$, which corresponds to 2 peaks of $\hat{\psi}$ per rotation of the binary secondary.}
\label{fig:MoIs2Angles}
\end{figure}

Next, the moments of inertia in the $\mathcal{G}$ frame are plotted in Fig.~\ref{fig:MoIs2} for the start of the simulation, highlighting the transient initialization effects on the shape in the first fraction of an orbit, and the longer-term settling of $I_{zz}$ and $I_{yy}$ due to spin-down. Such transience occupies $\sim$10\% of the first simulated orbit. Such transient periods are neglected for our calculations. Note that the $\mathcal{G}$ frame is not a principal axis frame: time-varying tidal distortion, especially concentrated towards the surface of the body, shifts the principal axis directions back and forth with respect to rocks in the deep interior. The 3-2-1 rotation from a principal axis frame, for the same period as Fig.~\ref{fig:MoIs2}, is given in Fig.~\ref{fig:MoIs2Angles}. Note the domination of the 3-2-1 angles by a semidiurnal signal in rotation about $z$, $\hat{\psi}$, which is due to such tidal effects. Such effects are more discernable isolating these plots to just a few orbits. 
Fig.~\ref{fig:Hconservation} shows the evolution of the total angular momentum, together with the planar spin angular momentum of the secondary and the orbital angular momentum. In principle a DEM conserves angular momentum; however, errors in contact resolution and time integration introduce small deviations, which is a known limitation in DEM models. Here the total angular momentum is well-conserved to within $1.5\times10^{-4}$ (relative) over the course of the full simulation, with smaller monotonic drift. Over the same interval, tidal torques transfer approximately $2\times10^{-3}$ of the total angular momentum from the spin angular momentum of the secondary to the orbital angular momentum, which is more than an order of magnitude larger than the numerical error. From the close agreement between computed and analytical approximations of the tidal torques (Fig.~\ref{fig:subfigB3}), we are confident that the spin evolution is driven by the tidal interaction, with a negligible contribution from numerical error.

\begin{figure}[h!]
\centering
\begin{subfigure}[b]{0.45\textwidth}
\centering
\includegraphics[width=2.6in]{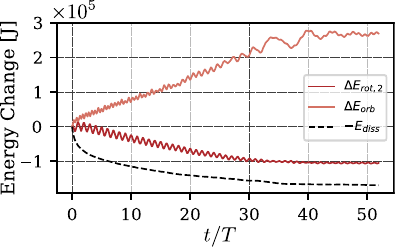} 
\caption{Change in total system energy, as computed by GRAINS, Eq.~\eqref{eq:diss} (dashed curve), compared to changes in orbital energy and the (approximate) rotational energy, Eq.~\eqref{Erot1}, of the secondary.}
\label{fig:Ec1A}
\end{subfigure} 
\hspace{0.08in}
\begin{subfigure}[b]{0.45\textwidth}
\centering
\includegraphics[width=3.1in]{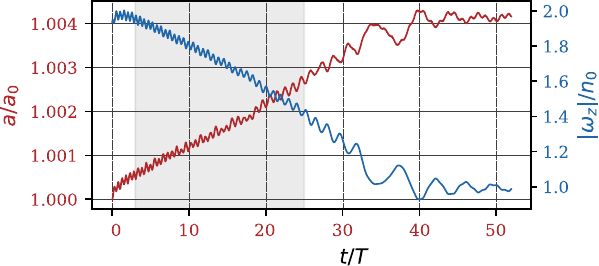} 
\caption{Semimajor axis vs. spin rate change over the simulation. The portion of the simulation used for $Q/k_{2}$ analysis is colored in gray, otherwise initial transience (towards $t/T\sim0$), or settling ($t/T\sim 30$ and higher).}
\label{fig:Ec1B}
\end{subfigure}
\caption{Change in system energy and secondary orbit/spin}
\label{fig:EnergyFigs1}
\end{figure}

Fig.~\ref{fig:Ec1A} displays the total system energy change from the start of the simulation (dashed line), as tracked by GRAINS, along with computed changes in the orbital energy of the secondary and its rotational energy (see Section 2.2.4). $E_{\text{diss}}$ relates to the dissipative power as
\begin{equation}
\label{eq:diss}
    \dot{E}_{\text{diss}} = -\frac{dE_{\text{mech}}}{dt} = -\frac{d}{dt}(E_{\text{kin, t}} + E_{\text{kin,rot}} + E_{\text{gp}} + E_{\text{el}}),
\end{equation}
where $E_{\text{kin, t}}$, $E_{\text{kin,r}}$, $E_{\text{gp}}$, and $E_{\text{el}}$ are the translational and rotational kinetic energy (summed over all particles), the pairwise gravitational potential energy (see Eq. \eqref{Nbody1}), and the stored elastic energy in each contact, respectively. The total energy dissipated amounts to $\sim$0.25\% of the total system energy. Much of the rotational energy lost from spin-down is not internally dissipated, but rather transferred to the orbit. We note that rotational energy of the secondary is a very tiny fraction of total system energy, which is dominated by the orbital energy. The spin-orbit energy exchange is most evident from Fig.~\ref{fig:Ec1B}, which shows the approximately inverse relationship between secondary semimajor axis $a$ and its rotation rate about the $z$ axis, $\omega_{z}$. This figure shades in gray the period of the simulation for which all $Q/k_{2}$ computations are performed, as earlier times are avoided for initial transience, and tidal locking occurs towards $t/T\sim 30$. 


\subsection{Case 2: Non-principal axis rotation}
\begin{figure}[h!]
\centering
\begin{subfigure}[b]{0.45\textwidth}
\centering
\includegraphics[width=3.1in]{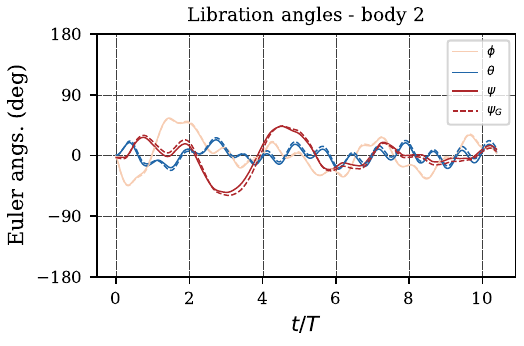}
\caption{Euler angles of the principal axis body frame $\mathcal{B}$ (solid) and topography frame $\mathcal{G}$ (dashed).}
\label{fig:subfigA1}
\end{subfigure} 
\hspace{0.27in}
\begin{subfigure}[b]{0.45\textwidth}
\centering
\includegraphics[width=3.1in]{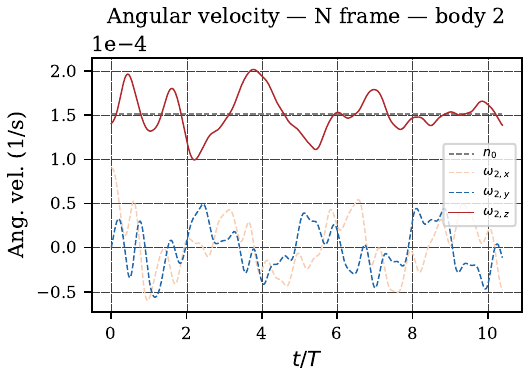}
\caption{Angular velocities of the body frame $\mathcal{B}$, with initial mean motion indicated by $n_{0}$.}
\label{fig:subfigA2}
\end{subfigure}
\vspace{0.2in} 
\begin{subfigure}[b]{0.45\textwidth}
\centering
\includegraphics[width=3.1in]{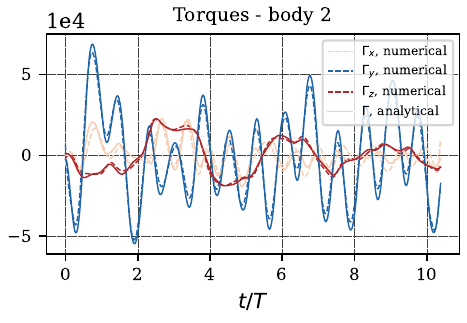}
\caption{Components of tidal torques on the rubble pile secondary. Numerically recovered (dashed) and analytical approximation (solid).}
\label{fig:subfigA3}
\end{subfigure} 
\hspace{0.27in} 
\begin{subfigure}[b]{0.45\textwidth}
\centering
\includegraphics[width=3.1in]{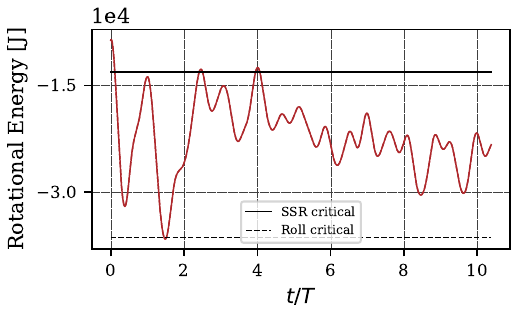}
\caption{Rotational energy of the secondary, as computed by Eq.~\eqref{Erot1}, for the case of non-principal axis rotation.}
\label{fig:subfigA4}
\end{subfigure}
\caption{Simulation results for Case 2 (NPA). $\mathcal{B}$ frame Euler angles (a) correspond to the rotation matrix $[\mathcal{BO}']$, whereas $\mathcal{G}$ corresponds to the rotation matrix $[\mathcal{GO}']$. The tidal torques (c) are provided in the $\mathcal{B}$ frame. Note the $x$ axis, roughly aligned with the direction towards the primary, experiences very small torque. This matches an expectation from \cite{Meyer_EnergyDissBinary}. Rotational energy (d) is insufficient to end up in a super-synchronous rotation case, as evidenced by $\mathcal{E} < \mathcal{E}_{\text{SSR}}$. Furthermore there is enough energy for roll,  $\mathcal{E} > \mathcal{E}_{\text{roll}}$. See the discussion following Eq.~\eqref{Erot1} for an explanation of the calculation, significance, and limitations of this rotational energy quantity $\mathcal{E}$.}
\label{fig:four_figures}
\end{figure}

Important simulation information is summarized in Table~\ref{table:caseInfo}.  
The rubble pile secondary is initialized in a spin state with some significant initial nonzero yaw and roll, though not enough to initialize a super-synchronous rotation state. The resulting attitude evolution is given in terms of Euler angles in Fig.~\ref{fig:subfigA1} and in terms of the angular velocity of the $\mathcal{B}$ frame with respect to $\mathcal{O}'$ in Fig.~\ref{fig:subfigA2}. 
At this spin energy level of $\mathcal{E}_{\text{roll}}<\mathcal{E}<\mathcal{E}_{\text{SSR}}$, it is possible for the body to end up in a ``barrel instability", first noted by \cite{cuk_2021}, though such a condition was not observed in this particular case. 
Fig.~\ref{fig:subfigA3} gives the tidal torque about the $x,y,z$ body axes of the binary secondary, as reproduced from the outputs of GRAINS.  Strong agreement can be noted between the numerically recovered estimate of the tidal torque (dashed lines), and the computation based on MacCullagh's approximation (solid lines). Torque along the long axis ($x$) is a little bit lower than the other two body axes, which is expected for a bounded rotation state. 
Fig.~\ref{fig:subfigA4} gives the rotational energy of the secondary, as computed by Eq.~\ref{Erot1}, for the case of non-principal axis rotation. The energy is 
not enough to end up in a supersynchronous rotation state, because $\mathcal{E} < \mathcal{E}_{\text{SSR}}$. This energy quantity, while not exact (See Section 2.2.4), nonetheless successfully predicted the transition from supersynchronous to bounded spin state in Case 1. 

\subsection{Tidal Computations}
\subsubsection{Estimating $Q/k_{2}$}
\begin{figure}[h!]
\centering
\begin{subfigure}[b]{0.45\textwidth}
\centering
\includegraphics[width=3.1in]{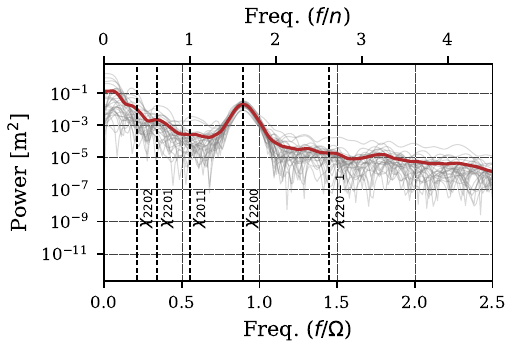}
\caption{Case 1 - Super Synchronous Rotation.}
\label{fig:RockModes1}
\end{subfigure} 
\hspace{0.27in}
\begin{subfigure}[b]{0.45\textwidth}
\centering
\includegraphics[width=3.1in]{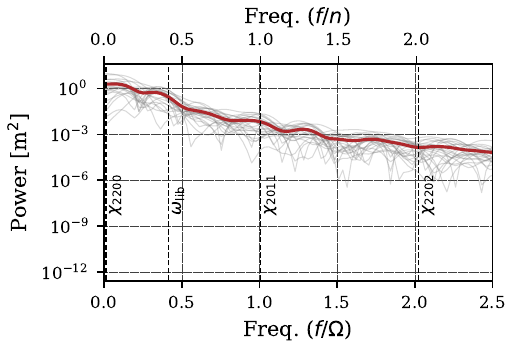}
\caption{Case 2 - Non-principal Axis Rotation.}
\label{fig:RockModes2}
\end{subfigure}
\caption{Spectral decomposition of the surface movement. The gray lines show the movement of individual particles at the surface, and at which frequency they are excited. The red line shows the average response of all the particles, with the dashed vertical lines showing the system frequencies given by Eq. \eqref{eq:chi}.}
\label{fig:RockModes}
\end{figure}

\begin{figure}[h!]
    \centering
    \includegraphics[]{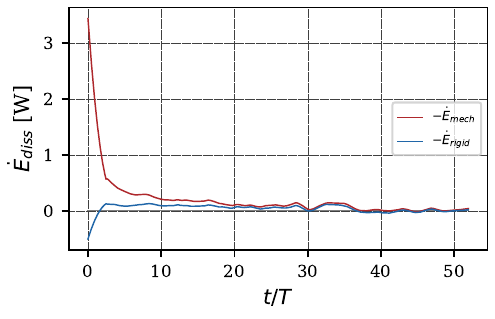}
    \caption{The amount of dissipation calculated from the total GRAINS energy Eqs. \eqref{eq:diss} and the change in rigid-body energy Eq. \eqref{TotalEApp}.}
    \label{fig:diss_calculation}
\end{figure}

\begin{figure}[h!]
\centering
\includegraphics[]{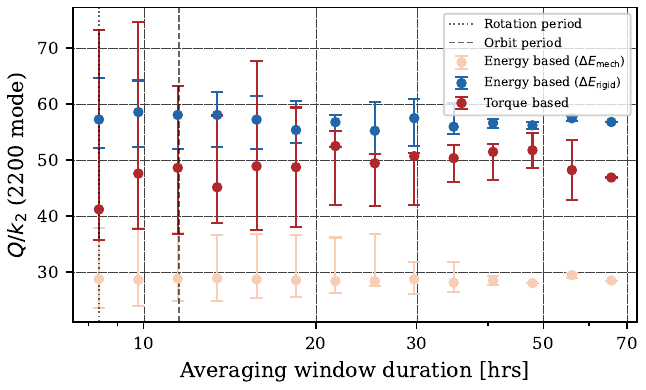}
\caption{$Q/k_{2}$ as computed from Eq.~\eqref{eq:matrix-l2} for the 2200 mode only. Computations using energy rate and torque-averaging approaches agree to well within an order of magnitude, despite computation via different approaches. Notably, the two energy-based approaches bound the torque-based results.}
\label{fig:Qk2ForReal1}
\end{figure}

\begin{figure}[h!]
\centering
\includegraphics[]{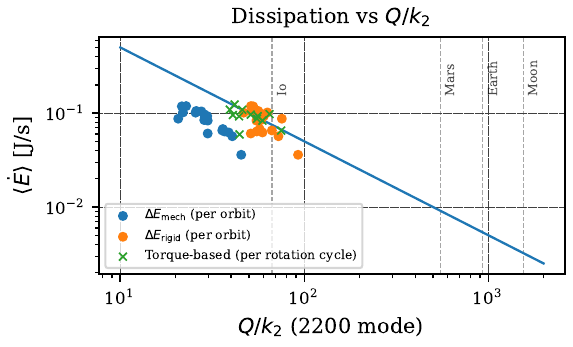}
\caption{Dissipativity of modeled rubble pile binary secondary in the context of other celestial bodies. The slope is the theoretical value calculated using Eq. \eqref{eq:matrix-l2} using the average $Q/k_{2}$. The values for the other celestial bodies come from \cite{Ray2001} (Earth), \cite{Williams2015} (Moon), \cite{Brozovic2025} (Mars), and \cite{Lainey2009} (Io).}
\label{fig:Qk2_slope}
\end{figure}

Given the successful recovery of a tidal torque signal, the equations outlined in Section 2.2 can be applied to estimate $Q/k_{2}$ and other parameters of interest for the case of super-synchronous rotation, Case 1, and also for the other cases. Because GRAINS tracks the locations of individual mass elements of the rubble pile, we can analyze the radial displacement of the mass elements, and the associated frequency of the motion, to characterize the spectral response of the rubble pile. For Case 1, Fig.~\ref{fig:RockModes1} gives a spectral decomposition of the surface movement using a discrete Fourier transform of the radius of individual particles over time (computed between orbit 5 and 25 to focus on the SSR period of the simulation), with individual spectra shown in gray, and the average response as a red curve. There is a clear peak at $\chi_{2200}$, the semidiurnal mode. This is due to tidal response of the secondary to the primary. For this mode, Fig.~\ref{fig:Qk2ForReal1} estimates $Q/k_{2}$ using both the torque- and energy-based methods. For case 2, shown in Fig.~\ref{fig:RockModes2}, the fact that it is close to spin-orbit locking makes the $\chi_{2200}$ equal to zero and thus other modes become important. Orbital modes (i.e. frequencies around the orbital period) like $\chi_{2011}$ can become more important if the eccentricity is high. However, it can be seen here that for this case there is no clear single mode that causes tidal dissipation, impeding a straightforward computation of $Q/k_{2}$. Future work will look at trying to find effective $Q/k_{2}$ that combine different modes, which requires adaptation of the energy and torque calculations. Another type of mode that becomes more important is the longitudinal libration normal mode, given by \citep{Rambaux2012},
\begin{equation}
    \omega_{\text{lib}}= n\sqrt{3\frac{B-A}{C}},
\end{equation}
which can be a source of dissipation as well \citep{Efroimsky2018}. As seen before in Fig.~\ref{fig:MoIs2}, due to the initial transient and the continuous adaption of the global shape of Dimorphos because of the dynamical conditions and changing spin over time, the dissipation rate calculated by the total change in mechanical energy, Eq. \eqref{eq:diss}, might overestimate the tidal contribution. Therefore, we also present the dissipation as purely coming from the rigid-body energy given in Eq. \eqref{TotalEApp}. The dissipation coming from these two different approaches can be seen in Fig \ref{fig:diss_calculation}. The rigid-body dissipation during the initial transient sees much less dissipation, and during the SSR period it measures a slightly lower rate as it does not consider the global shape change due to the slowing down of the spin. This approach can still miss some dissipation due to shape changes happening due to the tides, and thus it is argued that the true tidal dissipation lies somewhere between the mechanical and rigid-body derived values. This zone is also exactly where the results from the torque-based approach lie, supporting our computational approach. The results in Fig. ~\ref{fig:Qk2ForReal1} agree for the two different energy-based approaches and the torque-based approach to well within an order of magnitude. Overall, this is a rather dissipative finding for the simulated binary secondary, and Fig.~\ref{fig:Qk2_slope} contextualizes the $Q/k_{2}$ estimates in the context of other celestial bodies in the solar system.


\section{Discussion}
We show clearly that the semidiurnal tidal mode is present in our simulation of a rubble pile binary secondary. Conducting a study of spin-down from a planar super-synchronous spin state, we compute $Q/k_{2} \sim 20-80$ over all approaches and all averaging intervals, with some explained discrepancy if a torque- or energy-based approach is used. This is a rather dissipative result, but to recover a predictive value necessitates further study, focused on the poorly constrained mechanical properties of the constituent materials, their impact on the estimate, and the impact of various modeling strategies and assumptions common in DEM studies. As a point of comparison to our $Q/k_{2}$ calculation, \cite{PouNimmo2023} give estimated ranges of $10^{2} < Q_{s}/k_{2,s} < 10^{6}$ for about a dozen binary secondaries. That work makes use of observational studies of implied historical orbital evolution (for estimating $Q_{p}/k_{2,p}$ for the primary) or implied historical spin evolution (for estimating $Q_{s}/k_{2,s}$ for the secondary). 
Such observational studies are a vital anchor for numerical physics -- providing real constraints that the emulated systems should be tuned to respect. However, as most binary secondaries are already in a settled state, there is a dearth of direct measures of dissipativity of binary secondaries. Properly tuned numerical models can give some estimates for quantities of interest such as $Q/k_{2}$ for a given system/scenario, but much more than this, have the potential for providing great insights into tidal dissipation in such systems, which is poorly understood. Although rubble-pile asteroids are expected to exhibit viscoelastic behavior at the macroscopic scale, the effective rheology remains poorly understood and may differ substantially from the conventional viscoelastic models used to describe planetary bodies. We see a path forward for addressing this, by holistically studying the causal chain from the underlying governing granular mechanical properties to emergent dissipative behavior. Such insights would be comparatively difficult to gather from observational studies, where we are limited to surface-level studies of mainly settled systems. 

We also remind the reader of the recent work by \cite{DellaGiustina24}. Our setup is quite different, with a larger particle budget, the use of non-spherical mass elements, their split of mass elements across both bodies vs. our concentration of all mass elements into the secondary, and their focus on the dissipation within the primary vs. ours on rotational settling of the secondary. We also note that in comparison to our results, their orbit radius seems more regular, perhaps due to the less nonspherical secondary. In contrast to our quality factor estimates roughly estimated from the change in rotational energy $\mathcal{E}$, they apply the formula $Q_{\text{tide}} = \left(2\Delta t|\omega_{1} - n|\right)^{-1}$ \citep{EfroimskyAgain} for primary rotation rate $\omega_{1}$ and orbital mean motion $n$, and temporal lag $\Delta t$. Their result of $Q_{\text{tide}} = 8.97$, for the primary, was an order of magnitude smaller than prior estimates for monoliths \citep{GoldreichSari}. This agrees with our observation that our numerically simulated rubble piles are highly dissipative.

The possibility of binary asteroid systems being on the more dissipative side of prior estimates in literature, one hypothesis stemming from the tendency of \textit{numerical models} to be rather dissipative, would have significant implications for the scientific community. On the eve of the upcoming Hera mission to study the aftermath of the DART impact on Dimorphos, we may not have to wait long for helpful observational data. \cite{Dimorphos_LightCurve} found, via lightcurve analysis, that the DART mission likely put Dimorphos in a fairly excited attitude state, with large (though probably bounded) yaw angles, and the possibility for unbounded roll, though poorly constrained by the lightcurve analysis. The final state observed by Hera could provide some constraints on the dissipativity of Dimorphos. 
Further study with high-fidelity modeling is needed to determine the nature of the disagreement between numerical results and the estimates of $Q/k_{2}$ obtained by means of scaling analysis and astrometry. There are also various inconsistencies within the literature. \cite{NimmoMatsuyama} argues for shallow friction-dominated dissipation, whereas \cite{GoldreichSari} argues for dissipation dominated by yielding at sharp points on rocks in the asteroid interior. In the viewpoint of the latter, effective rigidity becomes dynamically important for dissipation, whereas \cite{Efroimsky_binaries} instead argues that effective viscosity is much more important than effective rigidity. Such theoretical disputes can and should be resolved by numerical study. 

The rubble pile secondary, when initialized, exhibits a brief transient response in the start of the simulation as it settles towards a more equilibrium shape. This is notable from the short spike in the very start of Fig.~\ref{fig:MoIs2}. We find that this happens rather quickly, in agreement with the findings of \cite{Tange_nonhydrostatic}, which by contrast was conducted in pkdgrav with smooth spheres. This issue is just one of many implementational details that we discuss for conducting scientific studies of tidal dissipation in discrete element models. We note also that the high dissipativity of our simulated rubble piles is in agreement with a finding of \cite{Wimarsson_GRAINS24} that some of the aggregating ``moonlets" in binary asteroid systems tidally lock extremely quickly, in a timespan of days. In a follow-up work, the authors further show that the efficiency of the tidal locking mechanism strongly varies with the distance at which a moon has formed, emphasizing that a change in semi-major axis of merely one per cent can determine whether a body remains in an SSR state \citep{Wimarsson_25}. These studies were also conducted in GRAINS. 

For longer-term numerical studies, our suggested workflow is to carefully and rigorously combine high- and low-fidelity (e.g. non-$N$ body) models. Relatively short-duration high-fidelity simulations, such as the ones conducted in this work, can be used to provide estimated tidal parameters for scenarios of interest based on orbit/body sizes, body shape, and spin states. Then, a faster simulation architecture, modeling only the rigid-body physics \citep{Davis_GUBAS}, plus various parametric equations for YORP and tides, is loaded with an approximate tidal torque which is adjusted as needed based on the spin/orbit regime. See e.g. \cite{Cueva_2024} and \cite{Meyer_EnergyDissBinary} for recent relevant studies. It is worth mentioning again that changes to the unimodal model assumptions, beyond the ubiquitous MacDonald case (e.g., \cite{EfroimskyTidal}), can produce quite different outcomes in long-term numerical simulations. This was observed, for example, in a recent study with long-term numerical simulations of the tidal settling of Europa \citep{BurnettHayne_Icarus2}. Given the distinct bulk rheological nature of rubble piles, one could argue in favor of the development of distinct low-fidelity tidal models specifically for rubble pile asteroids, which are designed to serve as valid first-order approximations of the high-fidelity physics as revealed by GRAINS or alternate $N$-body models. 

\section{Conclusions}
In this work we outline numerical recipes for conducting detailed studies of tidal dissipation in $N$-body models of rubble pile binary systems, and apply the result to a system similar in size, scale, and geometry to Didymos-Dimorphos. Focusing on tidal dissipation in the secondary, we render the primary as a point mass so that all remaining particles in the $N$-body model can be used to build the secondary. Key results from this study include the successful emergence and identification of tidal distortion, tidal lag, and resulting dissipation. This is the first study to our knowledge to estimate the resulting dissipativity of the rubble pile binary secondary from an $N$-body model. We numerically estimate a highly dissipative result of $Q/k_{2} \sim 20-80$ for a Dimorphos-like binary, which is on/below the low end for binary secondaries predicted by various other studies. We note that the effect of physical properties of the constitutive particles on dissipativity should be explored, and also that the value of $Q/k_{2}$ will depend on the dynamical scenario. More work is needed to determine if this dissipative result is a consequence of modeling assumptions or representative of real small bodies.

We introduce a number of helpful strategies needed for conducting further numerical studies. These include the definition of a novel ``topography" frame, fixed with respect to select rocks deep in the interior, the identification and exclusion of initial transient behaviors inherent to numerical rubble pile models. We also compile helpful insights from literature related to DEM modeling of granular material, tidal theory, and important details of binary asteroid morphology and dynamics as revealed by past studies and missions. Lastly, we define numerical computations, based in classical theory, for estimating various tidal quantities of interest. Due to fundamentally different physics governing dissipation in rubble pile asteroids, in bulk these bodies need not obey classical tidal equations founded on certain rheological assumptions. Instead, we show that a discrete-element simulation can be used as a numerical laboratory for studying asteroid and granular body tidal physics without presupposing a bulk rheological model. 
We advocate for supplementary studies relying on numerical (and also in-laboratory) experiments, focused on the goal of improving our fundamental understanding of the underlying physics of dissipation in granular media in low-gravity settings. 

\section{Funding Acknowledgments} 
Funded/co-funded by the European Union (Horizon Europe, FAAST-MSCA, 101063274, SEISMOR-MSCA, 101264707; ERC, TRACES, 101077758). Views and opinions expressed are however those of the authors only and do not necessarily reflect those of the European Union, the European Research Executive Agency, or the European Research Council. Neither the European Union nor the granting authority can be held responsible for them. This work was also partly carried out within the Space It Up! project funded by the Italian Space Agency (ASI) and the Italian Ministry of University and Research (MUR) under contract No. 2024-5-E.0, CUP No. I53D24000060005.

\bibliographystyle{apacite}
\bibliography{references.bib}

\end{document}